\documentclass[12pt]{article}%
\usepackage{amsmath}
\usepackage{amsfonts}
\usepackage{amssymb}
\usepackage{graphicx}
\usepackage{cite}
\allowdisplaybreaks
\begin{document}

\author{Hartmut Wachter\thanks{e-mail: Hartmut.Wachter@physik.uni-muenchen.de}\\\hspace{0.4in}\\Max-Planck-Institute\\for Mathematics in the Sciences\\Inselstr. 22, D-04103 Leipzig, Germany\hspace{0.16in}\\\hspace{0.4in}\\Arnold-Sommerfeld-Center\\Ludwig-Maximilians-Universit\"{a}t\\Theresienstr. 37, D-80333 M\"{u}nchen, Germany}
\title{q-Translations on quantum spaces}
\date{}
\maketitle

\begin{abstract}
Attention is focused on quantum spaces of particular importance in physics,
i.e. two-dimensional quantum plane, q-deformed Euclidean space in three or
four dimensions, and q-deformed Minkowski space. Each of these quantum spaces
can be combined with its symmetry algebra to form a Hopf algebra. The Hopf
structures on quantum space coordinates imply their translation. This article
is devoted to the question how to calculate translations on the quantum spaces
under consideration.
\end{abstract}

\section{Introduction}

\subsection{General motivation for deforming spacetime}

Relativistic quantum field theory is not a fundamental theory, since its
formalism leads to divergencies. In some cases like that of quantum
electrodynamics one is able to overcome the difficulties with the divergencies
by applying the so-called renormalization procedure due to Feynman, Schwinger,
and Tomonaga \cite{Schw}. Unfortunately, this procedure is not successful if
we want to deal with quantum gravity. Despite the fact that gravitation is a
rather weak interaction we are not able to treat it perturbatively. The reason
for this lies in the fact that transition amplitudes of n-th order to the
gravitation constant diverge like a momentum integral of the general form
\cite{Wei}
\begin{equation}
\int p^{2n-1}dp. \label{ImIntN}%
\end{equation}
This leaves us with an infinite number of ultraviolet divergent Feynman
diagrams that cannot be removed by redefining finitely many physical parameters.

It is surely legitimate to ask for the reason for these fundamental
difficulties. It is commonplace that the problems with the divergences in
relativistic quantum field theory result from an incomplete description of
spacetime at very small distances \cite{Schw}. Niels Bohr and Werner
Heisenberg have been the first who suggested that quantum field theories
should be formulated on a spacetime lattice \cite{Cass, Heis}. Such a
spacetime lattice would imply the existence of a smallest distance $a$ with
the consequence that plane-waves of wave-length smaller than twice the lattice
spacing could not propagate. In accordance with the relationship between
wave-length $\lambda$ and momentum $p$ of a plane-wave, i.e.
\begin{equation}
\lambda\geq\lambda_{\min}=2a\quad\Rightarrow\quad\frac{1}{\lambda}\sim p\leq
p_{\max}\sim\frac{1}{2a},
\end{equation}
it follows then that physical momentum space would be bounded. Hence, the
domain of all momentum integrals in Eq. (\ref{ImIntN}) would be bounded as
well with the consequence that momentum integrals should take on finite values.

\subsection{q-Deformation of symmetries as an attempt to get a more detailed
description of nature}

Discrete spacetime structures in general do not respect classical Poincar\'{e}
symmetry. A possible way out of this difficulty is to modify not only
spacetime but also its corresponding symmetries. How are we to accomplish
this? First of all let us recall that classical spacetime symmetries are
usually described by Lie groups. If we realize that Lie groups are manifolds
the Gelfand-Naimark\textsf{\ }theorem tells us that Lie groups can be
naturally embedded in the category of algebras\textsf{\ }\cite{GeNe}. The
utility of this interrelation lies in formulating the geometrical structure of
Lie groups in terms of a Hopf structure \cite{Hopf}. The point is that during
the last two decades generic methods have been discovered for continuously
deforming matrix groups and Lie algebras within the category of Hopf algebras.
It is this development which finally led to the arrival of quantum groups and
quantum spaces \cite{Ku83, Wor87, Dri85, Jim85, Drin86, RFT90, Tak90}.

From a physical point of view the most realistic and interesting deformations
are given by q-deformed versions of Minkowski space and Euclidean spaces as
well as their corresponding symmetries, i.e. respectively Lorentz symmetry and
rotational symmetry \cite{CSSW90, PW90, SWZ91, Maj91, LWW97}. Further studies
even allowed to establish differential calculi on these q-deformed quantum
spaces \cite{WZ91, CSW91, Song92, OSWZ92} representing nothing other than
q-analogs of translational symmetry. In this sense we can say that
q-deformations of the complete Euclidean and Poincar\'{e} symmetries are now
available \cite{Maj93-2}. Finally, Julius Wess and his coworkers were able to
show that q-deformation of spaces and symmetries can indeed lead to the wanted
discretizations of the spectra of spacetime observables \cite{FLW96, Wes97,
CHMW99, CW98}. This observation nourishes the hope that q-deformation might
give a new method to regularize quantum field theories \cite{GKP96, MajReg,
Oec99, Blo03}.

\subsection{Intention and content of this work}

In order to formulate quantum field theories on quantum spaces it is necessary
to provide us with some essential tools of a q-deformed analysis
\cite{Wess00}. The main question is how to define these new tools, which
should be q-analogs of classical notions. Towards this end the considerations
of Shahn Majid have proved very useful \cite{Maj91Kat, Maj94Kat, Maj93-Int,
Maj94-10}. The key idea of this approach is that all the quantum spaces to a
given quantum symmetry form a braided tensor category. Consequently,
operations and objects concerning quantum spaces must rely on this framework
of a braided tensor category, in order to guarantee their well-defined
behavior under quantum group transformations. This so-called principle of
covariance can be seen as the essential guideline for constructing a
consistent theory.

In our previous work \cite{WW01, BW01, Wac02, Wac03, MSW04, SW04} we have
applied these general considerations, exposed in Refs. \cite{Maj94-10,
Maj93-Int, Maj93-6, Maj93-5, Maj95}, to quantum spaces of physical importance,
i.e. q-deformed quantum plane, q-deformed Euclidean space with three or four
dimensions, and q-deformed Minkowski space. In this manner, we obtained
explicit expressions for computing star products, operator representations of
partial derivatives and symmetry generators, q-integrals, and q-exponentials.
Since the fundamental laws of physics should be invariant under translations
in space and time, we would like to include q-deformed translations into our
framework. In this respect, the main concern of this article is to discuss
methods for calculating q-deformed translations on the quantum spaces under consideration.

In particular, we intend to proceed as follows. In Sec. \ref{BasSec} we
briefly describe the basic ideas our investigations are based on. For further
details we recommend Refs.\ \cite{Maj95} and \cite{ChDe96}. In the subsequent
sections we apply these reasonings to quantum spaces of physical importance,
i.e. the two-dimensional quantum plane, the q-deformed Euclidean space with
three or four dimensions, and the q-deformed Minkowski space. In doing so, we
get formulae which can be viewed as q-analogs of Taylor rules. A conclusion
closes our investigations by providing the reader with interesting remarks
about the relationship between q-translations and the other objects of
q-analysis. For reference and for the purpose of introducing consistent and
convenient notation, we give a review of key notation and results in the
Appendices \ref{AppNot} and \ref{AppQuan}. Appendix \ref{AppTrans} is devoted
to coordinate transformations on q-deformed Minkowski space. It was added for
the purpose of being able to combine the results in this article with those
from our previous work.

\section{Basic ideas about q-deformed translations\label{BasSec}}

First of all, it should be emphasized that q-deformed translations are an
essential ingredient of q-analysis. As already mentioned, q-analysis is
formulated within the framework of quantum spaces \cite{Wess00, Maj94-10,
Maj93-6}. From a mathematical point of view, a quantum space is defined as
comodule of a quantum group \cite{Ku83, Wor87, Drin86, PW90}. For our purposes
it is at first sufficient to consider a quantum space as an algebra
$\mathcal{A}_{q}$ of formal power series in non-commuting coordinates
$X^{1},X^{2},\ldots,X^{n}$, i.e.%

\begin{equation}
\mathcal{A}_{q}=\mathbb{C}\left[  \left[  X^{1},\ldots,X^{n}\right]  \right]
/\mathcal{I},
\end{equation}
where $\mathcal{I}$ denotes the ideal generated by the relations of the
non-commuting coordinates. Notice that in Appendix \ref{AppQuan} we collected
the coordinate relations for the quantum spaces we are dealing with.

Physical theories have to make predictions about measurable quantities.
Normally, these predictions are given as real numbers. It is indeed a
non-trivial question how the measurable values arise in a physical theory
based on non-commutative spacetime structures. Usually, one seeks a complete
set of commuting observables. The measurable values are then given by the
corresponding eigenvalues. For the details of this approach we refer the
reader to Refs. \cite{CW98, Fio93, Zip95, FBM03}. In our work, however, we
follow a method inspired by deformation quantization \cite{BFF78, Moy49,
MSSW00}. Towards this end, we translate operations on non-commutative
coordinate algebras into those on commutative ones. The reason for this to be
possible lies in the fact that the non-commutative algebras\ we are dealing
with satisfy the \textit{Poincar\'{e}-Birkhoff-Witt property}. This property
implies that the monomials of a given normal ordering constitute a basis of
$\mathcal{A}_{q}$. Due to this fact, we can establish a vector space
isomorphism between $\mathcal{A}_{q}$ and a commutative algebra $\mathcal{A}$
generated by ordinary coordinates $x^{1},x^{2},\ldots,x^{n}$:
\begin{gather}
\mathcal{W}:\mathcal{A}\longrightarrow\mathcal{A}_{q},\nonumber\\
\mathcal{W}((x^{1})^{i_{1}}\ldots(x^{n})^{i_{n}})\equiv(X^{1})^{i_{1}}%
\ldots(X^{n})^{i_{n}}.\label{AlgIsoN}%
\end{gather}

The above vector space isomorphism can even be extended to an algebra
isomorphism by introducing a non-commutative product in $\mathcal{A}$, the
so-called \textit{star product}. This product is defined via the relation
\begin{equation}
\mathcal{W}(f\circledast g)=\mathcal{W}(f)\cdot\mathcal{W}(g), \label{StarDef}%
\end{equation}
being tantamount to%
\begin{equation}
f\circledast g=\mathcal{W}^{-1}\left(  \mathcal{W}\left(  f\right)
\cdot\mathcal{W}\left(  g\right)  \right)  ,
\end{equation}
where $f$ and $g$ are formal power series in $\mathcal{A}$. If there is an
algebra $\mathcal{H}$ which provides actions upon the non-commutative algebra
$\mathcal{A}_{q}$, we are also able to introduce actions upon the
corresponding commutative algebra $\mathcal{A}$ by means of the relations
\begin{align}
\mathcal{W}(h\triangleright f)  &  =h\triangleright\mathcal{W}(f),\text{\quad
}h\in\mathcal{H}\text{, }f\in\mathcal{A}\text{,}\nonumber\\
\mathcal{W}(f\triangleleft h)  &  =\mathcal{W}(f)\triangleleft h,
\label{defrepN}%
\end{align}
or%
\begin{align}
h\triangleright f  &  =\mathcal{W}^{-1}\left(  h\triangleright\mathcal{W}%
(f)\right)  ,\nonumber\\
f\triangleleft h  &  =\mathcal{W}^{-1}\left(  \mathcal{W}(f)\triangleleft
h\right)  . \label{DefParAcN}%
\end{align}

Our investigations require to deal with tensor products of quantum spaces. In
general multiplication on tensor products of quantum spaces cannot be
performed componentwise, since elements of distinct quantum spaces do not
commute. For this to become more clear let us first recall that the quantum
spaces under consideration are modules of \textit{quasitriangular Hopf
algebras} (in the following these Hopf algebras are denoted by $\mathcal{H}$).
Essentially for us is the fact that its coproduct can be twisted by an
invertible element $\mathcal{R=R}_{[1]}\otimes\mathcal{R}_{[2]}\in
\mathcal{H}\otimes\mathcal{H}$ or its transposed inverse $\tau(\mathcal{R}%
^{-1})=\mathcal{R}_{21}^{-1}=\mathcal{R}_{[2]}^{-1}\otimes\mathcal{R}%
_{[1]}^{-1}$. More formally, we have%
\begin{align}
\mathcal{R}(\Delta h)  &  =(\tau\circ\Delta h)\mathcal{R},\quad h\in
\mathcal{H},\nonumber\\
\mathcal{R}_{21}^{-1}(\Delta h)  &  =(\tau\circ\Delta h)\mathcal{R}_{21}^{-1},
\end{align}
where $\Delta$ and $\tau$ stand for the coproduct on $\mathcal{H}$ and the
transposition map, respectively.

Furthermore, we\ demand that the tensor product of quantum spaces has to be a
module of $\mathcal{H}$. This is the case if the action of $\mathcal{H}$ upon
tensor products is given by
\begin{equation}
h\triangleright(u\otimes v)=(h_{(1)}\triangleright u)\otimes(h_{(2)}%
\triangleright v),
\end{equation}
where the coproduct of $h$ is written in the so-called Sweedler notation. The
requirement for the tensor product to be a module of $\mathcal{H}$ implies
that multiplication on tensor products of quantum spaces is determined by%
\begin{equation}
(u_{1}\otimes v_{1})(u_{2}\otimes v_{2})=(u_{1}(\mathcal{R}_{[2]}%
\triangleright u_{2}))\otimes((\mathcal{R}_{[1]}\triangleright v_{1})v_{2}),
\end{equation}
or alternatively by%
\begin{equation}
(u_{1}\otimes v_{1})(u_{2}\otimes v_{2})=(u_{1}(\mathcal{R}_{[1]}%
^{-1}\triangleright u_{2}))\otimes((\mathcal{R}_{[2]}^{-1}\triangleright
v_{1})v_{2}).
\end{equation}
For the details we recommend Refs. \cite{Maj95} and \cite{ChDe96}.

From the explicit form of the tensor product of quantum spaces we can read off
commutation relations between elements of distinct quantum spaces.
For\ quantum space coordinates these relations become%
\begin{align}
(1\otimes X^{i})(Y^{j}\otimes1)  &  =(\mathcal{R}_{[2]}\triangleright
Y^{j})\otimes(\mathcal{R}_{[1]}\triangleright X^{i})\nonumber\\
&  =(Y^{j}\triangleleft\mathcal{R}_{[2]})\otimes(X^{i}\triangleleft
\mathcal{R}_{[1]})\nonumber\\
&  =k\,\hat{R}_{kl}^{ij}\,Y^{k}\otimes X^{l}, \label{VerRN}%
\end{align}
or%
\begin{align}
(1\otimes X^{i})(Y^{j}\otimes1)  &  =(\mathcal{R}_{[1]}^{-1}\triangleright
Y^{j})\otimes(\mathcal{R}_{[2]}^{-1}\triangleright X^{i})\nonumber\\
&  =(Y^{j}\triangleleft\mathcal{R}_{[1]}^{-1})\otimes(X^{i}\triangleleft
\mathcal{R}_{[2]}^{-1})\nonumber\\
&  =k^{-1}(\hat{R}^{-1})_{kl}^{ij}\,Y^{k}\otimes X^{l}, \label{VerRInN}%
\end{align}
where repeated indices are to be summed. Notice that $\hat{R}_{kl}^{ij}$ and
$(\hat{R}^{-1})_{kl}^{ij}$ denote vector representations of $\mathcal{R}$ and
$\mathcal{R}_{21}^{-1},$ respectively. To understand the second identity in
(\ref{VerRN}) and (\ref{VerRInN}), one has to realize that%
\begin{align}
(S^{-1}\otimes S^{-1})\circ\mathcal{R}  &  =\mathcal{R},\\
(S^{-1}\otimes S^{-1})\circ\mathcal{R}^{-1}  &  =\mathcal{R}^{-1},\nonumber
\end{align}
and
\begin{equation}
S^{-1}(h)\triangleright a=a\triangleleft h,\quad h\in\mathcal{H},\quad
a\in\mathcal{A}_{q},
\end{equation}
where $S^{-1}$ stands for\ the inverse of the antipode of $\mathcal{H}$.

As we will immediately see, it is convenient to formulate translations of
quantum space coordinates by using the so-called L-matrix. This L-matrix
provides an elegant way to describe how quantum space coordinates commute with
arbitrary elements of other quantum spaces. A short glance at (\ref{VerRN})
and (\ref{VerRInN}) should make it obvious that we need two L-matrices. These
L-matrices, in the following denoted by $\mathcal{L}$ and $\mathcal{\bar{L}}$,
have to be subject to the relations%
\begin{align}
(1\otimes X^{i})(a\otimes1)  &  =(\mathcal{R}_{[2]}\triangleright
a)\,(\mathcal{R}_{[1]}\triangleright X^{i})\nonumber\\
&  =\big ((\mathcal{\bar{L}}_{x}\mathcal{)}_{j}^{i}\triangleright
a\big )\,X^{j},\\
(1\otimes a)(X^{i}\otimes1)  &  =(X^{i}\triangleleft\mathcal{R}_{[2]}%
)\,(a\triangleleft\mathcal{R}_{[1]})\nonumber\\
&  =X^{j}\big (a\triangleleft(\mathcal{L}_{x})_{j}^{i}\big ),
\end{align}
and%
\begin{align}
(1\otimes X^{i})(a\otimes1)  &  =(\mathcal{R}_{[1]}^{-1}\triangleright
a)\,(\mathcal{R}_{[2]}^{-1}\triangleright X^{i})\nonumber\\
&  =\big ((\mathcal{L}_{x}\mathcal{)}_{j}^{i}\triangleright a\big )\,X^{j},\\
(1\otimes a)(X^{i}\otimes1)  &  =(X^{i}\triangleleft\mathcal{R}_{[1]}%
^{-1})\,(a\triangleleft\mathcal{R}_{[2]}^{-1})\nonumber\label{LGXN}\\
&  =X^{j}\big (a\triangleleft(\mathcal{\bar{L}}_{x})_{j}^{i}\big ).
\end{align}

It should be noticed that the entries of the L-matrices live in the Hopf
algebra\ $\mathcal{H}$ that describes the symmetry of the quantum spaces.
Since L-matrices are later used to write down translations of quantum space
coordinates we need an algebra which contains both the Hopf algebra
$\mathcal{H}$ and the quantum space $\mathcal{A}_{q}.$ For this reason, let us
make contact with the notion of a \textit{cross product algebra }\cite{KS97}.
It is well-known that we can combine a Hopf algebra $\mathcal{H}$ with its
representation space $\mathcal{A}_{q}$ to form a left cross product algebra
$\mathcal{A}_{q}\rtimes\mathcal{H}$ built on $\mathcal{A}_{q}\otimes
\mathcal{H}$ with product%
\begin{equation}
(a\otimes h)(b\otimes g)=a(h_{(1)}\triangleright b)\otimes h_{(2)}g,\quad
a,b\in\mathcal{A},\mathcal{\quad}h,g\in\mathcal{H}. \label{LefCrosPro}%
\end{equation}
There is also a right-handed version of this notion called a right cross
product algebra $\mathcal{H}\ltimes\mathcal{A}_{q}$ and built on
$\mathcal{H}\otimes\mathcal{A}_{q}$ with product%
\begin{equation}
(h\otimes a)(g\otimes b)=hg_{(1)}\otimes(a\triangleleft g_{(2)})b.
\label{RigCrosPro}%
\end{equation}

Important for us is the fact that on a cross product algebra we can assign two
Hopf structures to quantum space coordinates. More concretely, their
corresponding coproduct, antipode, and counit take the form \cite{OSWZ92,
Maj93-2}%
\begin{align}
\Delta_{\bar{L}}(X^{i})  &  =X^{i}\otimes1+(\mathcal{\bar{L}}_{x})_{j}%
^{i}\otimes X^{j},\nonumber\\
\Delta_{L}(X^{i})  &  =X^{i}\otimes1+(\mathcal{L}_{x})_{j}^{i}\otimes
X^{j},\label{HopfStrucN}\\[0.16in]
S_{\bar{L}}(X^{i})  &  =-S(\mathcal{\bar{L}}_{x})_{j}^{i}\,X^{j},\nonumber\\
S_{L}(X^{i})  &  =-S(\mathcal{L}_{x})_{j}^{i}\,X^{j},\label{SExplN}\\[0.16in]
\varepsilon_{\bar{L}}(X^{i})  &  =\varepsilon_{L}(X^{i})=0.
\end{align}

In addition to this, there are opposite Hopf structures related to the above
ones via%
\begin{equation}
\Delta_{\bar{R}/R}=\tau\circ\Delta_{\bar{L}/L},\qquad S_{\bar{R}/R}=S_{\bar
{L}/L}^{-1},\qquad\varepsilon_{\bar{R}/R}=\varepsilon_{\bar{L}/L},
\label{RightHopf}%
\end{equation}
where $\tau$ again\ denotes the usual transposition of tensor factors. Notice
that for the antipodes to the opposite Hopf structures it holds%
\begin{align}
S_{\bar{R}}(X^{i})  &  =-X^{j}S^{-1}(\mathcal{\bar{L}}_{x})_{j}^{i}%
,\nonumber\\
S_{R}(X^{i})  &  =-X^{j}S^{-1}(\mathcal{L}_{x})_{j}^{i}, \label{S-1ExplN}%
\end{align}
which is a direct consequence of the Hopf algebra axiom%
\begin{equation}
a_{(2)}S^{-1}(a_{(1)})=\varepsilon(a).
\end{equation}

An essential observation is that coproducts of coordinates imply their
translations \cite{Maj93-2, SW04, Maj93-5, Maj93-7, Mey95}. This can be seen
as follows. The coproduct $\Delta_{A}$ on coordinates is an algebra
homomorphism. If the coordinates constitute a module coalgebra then the
algebra structure of the coordinates $X^{i}$ is carried over to their
coproduct $\Delta_{A}(X^{i}).$ More formally, we have
\begin{equation}
\Delta_{A}(X^{i}X^{j})=\Delta_{A}(X^{i})\Delta_{A}(X^{j})\quad\text{and\quad
}\Delta_{A}(h\cdot X^{i})=\Delta(h)\cdot\Delta_{A}(X^{i}). \label{Trans}%
\end{equation}
Due to this fact we can think of (\ref{HopfStrucN}) as nothing other than a
q-deformed addition law for vector components.

To proceed any further we have to realize that our algebra morphism
$\mathcal{W}^{-1}$ can be extended to cross products by
\begin{align}
\mathcal{W}_{L}^{-1}  &  :\mathcal{A}_{q}\rtimes\mathcal{H}\longrightarrow
\mathcal{A},\nonumber\\
\mathcal{W}_{L}^{-1}((X^{1})^{i_{1}}\ldots(X^{n})^{i_{n}}\otimes h)  &
\equiv\mathcal{W}^{-1}((X^{1})^{i_{1}}\ldots(X^{n})^{i_{n}})\,\varepsilon(h),
\label{WRh}%
\end{align}
or
\begin{align}
\mathcal{W}_{R}^{-1}  &  :\mathcal{H}\ltimes\mathcal{A}_{q}\longrightarrow
\mathcal{A},\nonumber\\
\mathcal{W}_{R}^{-1}(h\otimes(X^{1})^{i_{1}}\ldots(X^{n})^{i_{n}})  &
\equiv\varepsilon(h)\,\mathcal{W}^{-1}((X^{1})^{i_{1}}\ldots(X^{n})^{i_{n}}),
\label{WRh-1}%
\end{align}
with $\varepsilon$ being the antipode of the Hopf algebra $\mathcal{H}.$ It
should be noted that in some sense these mappings are related to the process
of transmutation introduced in Ref. \cite{Maj93-8}.

With (\ref{WRh}) and (\ref{WRh-1}) at hand we are able to define\ q-deformed
translations on an algebra of commutative functions:%
\begin{align}
f(x^{i}\oplus_{L/\bar{L}}y^{j})  &  \equiv\big (\left(  \mathcal{W}_{L}%
^{-1}\otimes\mathcal{W}_{L}^{-1}\right)  \circ\Delta_{L/\bar{L}}%
\big )(\mathcal{W}(f)),\nonumber\\
f(x^{i}\oplus_{R/\bar{R}}y^{j})  &  \equiv\big (\left(  \mathcal{W}_{R}%
^{-1}\otimes\mathcal{W}_{R}^{-1}\right)  \circ\Delta_{R/\bar{R}}%
\big )(\mathcal{W}(f)),\label{DefCoProN}\\[0.16in]
f(\ominus_{L/\bar{L}}\,x^{i})  &  \equiv\big (\mathcal{W}_{R}^{-1}\circ
S_{L/\bar{L}}\big )(\mathcal{W}(f)),\nonumber\\
f(\ominus_{R/\bar{R}}\,x^{i})  &  \equiv\big (\mathcal{W}_{L}^{-1}\circ
S_{R/\bar{R}}\big )(\mathcal{W}(f)). \label{DefAntiN}%
\end{align}
In the sequel of this article we will derive explicit formulae for the
operations in (\ref{DefCoProN}) and (\ref{DefAntiN}). As already mentioned,
this will be done for quantum spaces of physical importance, i.e.
two-dimensional quantum plane, q-deformed Euclidean space with three or four
dimensions, and q-deformed Minkowski space.

\section{Two-dimensional quantum plane\label{2dimPl}}

We would like to start our considerations about q-deformed translations with
the two-dimensional quantum plane \cite{Man88, SS90} (for its definition see
Appendix \ref{AppQuan}). Due to its simplicity this case provides a good
opportunity to explain our reasonings in detail. As already mentioned, it is
our aim to derive operator expressions that enable us to calculate the objects
in (\ref{DefCoProN}) and (\ref{DefAntiN}) for arbitrary commutative functions.

To this end, we first consider a possible coproduct for the quantum space
coordinates $X^{1}$ and $X^{2}.$ Inserting the expressions for the entries of
the L-matrix, the second coproduct in (\ref{HopfStrucN}) becomes \cite{MSW04}
\begin{align}
\Delta_{L}(X^{1})=\,  &  X^{1}\otimes1+\Lambda^{3/4}\tau^{-1/4}\otimes
X^{1},\nonumber\\
\Delta_{L}(X^{2})=\,  &  X^{2}\otimes1+\Lambda^{3/4}\tau^{1/4}\otimes
X^{2}\nonumber\\
\,  &  -\,q\lambda\Lambda^{3/4}\tau^{-1/4}T^{+}\otimes X^{1},
\end{align}
where $\lambda\equiv q-q^{-1}.$ Notice that $T^{-}$ and $\tau$ denote
generators of the quantum algebra $U_{q}(su_{2}),$ whereas $\Lambda$ is a
scaling operator.

Next, we try to compute expressions for coproducts of normally ordered
monomials. In doing so, it is convenient to introduce\ left and right
coordinates by
\begin{equation}
X_{l}^{\alpha}\equiv X^{\alpha}\otimes1\quad\text{and\quad}X_{r}^{\alpha
}\equiv(\mathcal{L}_{x}\mathcal{)}_{\beta}^{\alpha}\otimes X^{\beta}.
\label{DefRLCoor}%
\end{equation}
Using this kind of abbreviation the coproduct of a normally ordered monomial
can be written as
\begin{align}
\Delta_{L}((X^{1})^{n_{1}}(X^{2})^{n_{2}})  &  =(\Delta_{L}(X^{1}))^{n_{1}%
}(\Delta_{L}(X^{2}))^{n_{2}}\nonumber\\
&  =(X_{l}^{1}+X_{r}^{1})^{n_{1}}(X_{l}^{2}+X_{r}^{2})^{n_{2}}.
\label{CoAll2dimN}%
\end{align}

Now, it is our task to bring the coordinates with lower index $l$ to the left
of the coordinates with lower index $r.$ For this to achieve, we have to know
the commutation relations between right and left coordinates. With the help of
the commutation relations between quantum space coordinates and the generators
of $U_{q}(su_{2})$, i.e. \cite{SWZ91}%
\begin{align}
T^{+}X^{1}  &  =qX^{1}T^{+}+q^{-1}X^{2},\nonumber\\
T^{+}X^{2}  &  =q^{-1}X^{2}T^{+},\label{RelTKoKanfN}\\[0.16in]
T^{-}X^{1}  &  =qX^{1}T^{-},\nonumber\\
T^{-}X^{2}  &  =q^{-1}X^{2}T^{-}+qX^{1},\\[0.16in]
\tau X^{1}  &  =q^{2}X^{1}\tau,\nonumber\\
\tau X^{2}  &  =q^{-2}X^{2}\tau,\\[0.16in]
\Lambda X^{\alpha}  &  =q^{-2}X^{\alpha}\Lambda,\quad\alpha=1,2,
\label{RelTKoKendNN}%
\end{align}
one can verify in a straightforward manner that the following identities
hold:
\begin{align}
X_{r}^{1}X_{l}^{1}  &  =q^{-2}X_{l}^{1}X_{r}^{1},\nonumber\\
X_{r}^{1}X_{l}^{2}  &  =q^{-1}X_{l}^{2}X_{r}^{1},\nonumber\\
X_{r}^{2}X_{l}^{2}  &  =q^{-2}X_{l}^{2}X_{r}^{2}.
\end{align}
Recalling the q-binomial theorem \cite{KS97, Koo96, Kac00} these relations
imply the formulae%
\begin{align}
\Delta_{L}(X^{1})^{n_{1}}  &  =\sum_{k=0}^{n_{1}}%
\genfrac{[}{]}{0pt}{}{n_{1}}{k}%
_{q^{-2}}(X_{l}^{1})^{k}(X_{r}^{1})^{n_{1}-k},\nonumber\\
\Delta_{L}(X^{2})^{n_{2}}  &  =\sum_{k=0}^{n_{2}}%
\genfrac{[}{]}{0pt}{}{n_{2}}{k}%
_{q^{-2}}(X_{l}^{2})^{k}(X_{r}^{2})^{n_{2}-k}. \label{CoX22dim}%
\end{align}
Notice that the definition of the q-binomial coefficients is given in Appendix
\ref{AppNot}. Inserting the result of (\ref{CoX22dim}) into
Eqn.(\ref{CoAll2dimN}) finally\ yields
\begin{align}
\Delta_{L}((X^{1})^{n_{1}}(X^{2})^{n_{2}})  &  =\sum_{i_{1}=0}^{n_{1}}%
\sum_{i_{2}=0}^{n_{2}}q^{-(n_{1}-\,i_{1})i_{2}}%
\genfrac{[}{]}{0pt}{}{n_{1}}{i_{1}}%
_{q^{-2}}%
\genfrac{[}{]}{0pt}{}{n_{2}}{i_{2}}%
_{q^{-2}}\nonumber\\
&  \qquad\times(X_{l}^{1})^{i_{1}}(X_{l}^{2})^{i_{2}}(X_{r}^{1})^{n_{1}%
-\,i_{1}}(X_{r}^{2})^{n_{2}-i_{2}}, \label{BinFor2dimN}%
\end{align}
where we made use of%
\begin{equation}
(X_{r}^{1})^{n_{1}-k}(X_{l}^{2})^{l}=q^{-(n_{2}-k)l}(X_{l}^{2})^{l}(X_{r}%
^{1})^{n_{1}-k}.
\end{equation}
A short glance at (\ref{DefCoProN}) shows us that it remains to apply
$\mathcal{W}_{L}^{-1}\otimes\mathcal{W}_{L}^{-1}$ to expression
(\ref{BinFor2dimN}). Thus, we perform the following calculation:%
\begin{align}
&  \big (\mathcal{W}_{L}^{-1}\otimes\mathcal{W}_{L}^{-1}\big )\big ((X_{l}%
^{1})^{i_{1}}(X_{l}^{2})^{i_{2}}(X_{r}^{1})^{n_{1}-\,i_{1}}(X_{r}^{2}%
)^{n_{2}-i_{2}}\big )\nonumber\\
=\,  &  \big (\mathcal{W}_{L}^{-1}\otimes\mathcal{W}_{L}^{-1}%
\big )\big ((X^{1})^{i_{1}}(X^{2})^{i_{2}}(\mathcal{L}_{x}\mathcal{)}%
_{\alpha_{1}}^{1}(\mathcal{L}_{x}\mathcal{)}_{\alpha_{2}}^{1}\ldots
(\mathcal{L}_{x}\mathcal{)}_{\alpha_{n_{1}-\,i_{1}}}^{1}\nonumber\\
\,  &  \qquad\times(\mathcal{L}_{x}\mathcal{)}_{\beta_{1}}^{2}(\mathcal{L}%
_{x}\mathcal{)}_{\beta_{2}}^{2}\ldots(\mathcal{L}_{x}\mathcal{)}_{\beta
_{n_{2}-i_{2}}}^{2}\nonumber\\
\,  &  \qquad\otimes X^{\alpha_{1}}X^{\alpha_{2}}\ldots X^{\alpha
_{n_{1}-\,i_{1}}}X^{\beta_{1}}X^{\beta_{2}}\ldots X^{\beta_{n_{2}-i_{2}}%
}\big )\nonumber\\
=\,  &  \mathcal{W}_{L}^{-1}\big ((X^{1})^{i_{1}}(X^{2})^{i_{2}}%
(\mathcal{L}_{x}\mathcal{)}_{\alpha_{1}}^{1}(\mathcal{L}_{x}\mathcal{)}%
_{\alpha_{2}}^{1}\ldots(\mathcal{L}_{x}\mathcal{)}_{\alpha_{n_{1}-\,i_{1}}%
}^{1}\nonumber\\
\,  &  \qquad\times(\mathcal{L}_{x}\mathcal{)}_{\beta_{1}}^{2}(\mathcal{L}%
_{x}\mathcal{)}_{\beta_{2}}^{2}\ldots(\mathcal{L}_{x}\mathcal{)}_{\beta
_{n_{2}-i_{2}}}^{2}\big )\nonumber\\
\,  &  \otimes\mathcal{W}_{L}^{-1}\big (X^{\alpha_{1}}X^{\alpha_{2}}\ldots
X^{\alpha_{n_{1}-i_{1}}}X^{\beta_{1}}X^{\beta_{2}}\ldots X^{\beta_{n_{2}%
-i_{2}}}\big )\nonumber\\
=\,  &  (x^{1})^{i_{1}}(x^{2})^{i_{2}}\,\varepsilon\big ((\mathcal{L}%
_{x}\mathcal{)}_{\alpha_{1}}^{1}\ldots(\mathcal{L}_{x}\mathcal{)}%
_{\alpha_{n_{2}-i_{2}}}^{2}\big )\nonumber\\
\,  &  \otimes\mathcal{W}_{L}^{-1}\big (X^{\alpha_{1}}X^{\alpha_{2}}\ldots
X^{\alpha_{n_{1}-\,i_{1}}}X^{\beta_{1}}X^{\beta_{2}}\ldots X^{\beta
_{n_{2}-i_{2}}}\big )\nonumber\\
=\,  &  (x^{1})^{i_{1}}(x^{2})^{i_{2}}\,\delta_{\alpha_{1}}^{1}\delta
_{\alpha_{2}}^{1}\ldots\delta_{\alpha_{n_{1}-\,i_{1}}}^{1}\delta_{\beta_{1}%
}^{2}\delta_{\beta_{2}}^{2}\ldots\delta_{\beta_{n_{1}-\,i_{1}}}^{2}\nonumber\\
\,  &  \otimes\mathcal{W}_{L}^{-1}\big (X^{\alpha_{1}}X^{\alpha_{2}}\ldots
X^{\alpha_{n_{1}-\,i_{1}}}X^{\beta_{1}}X^{\beta_{2}}\ldots X^{\beta
_{n_{2}-i_{2}}}\big )\nonumber\\
=\,  &  (x^{1})^{i_{1}}(x^{2})^{i_{2}}\otimes\mathcal{W}_{L}^{-1}%
\big ((X^{1})^{n_{1}-\,i_{1}}(X^{2})^{n_{2}-i_{2}}\big )\nonumber\\
=\,  &  (x^{1})^{i_{1}}(x^{2})^{i_{2}}\otimes(x^{1})^{n_{1}-\,i_{1}}%
(x^{2})^{n_{2}-i_{2}}. \label{UmRech}%
\end{align}
Notice that the fourth equality follows from $\varepsilon(\mathcal{L}_{\beta
}^{\alpha}\mathcal{)}=\delta_{\beta}^{\alpha}$ and the algebra isomorphism
(\ref{AlgIsoN}) is fixed by%
\begin{equation}
\mathcal{W}\left(  (x^{1})^{n_{1}}(x^{2})^{n_{2}}\right)  \equiv(X^{1}%
)^{n_{1}}(X^{2})^{n_{2}}. \label{OpRep1}%
\end{equation}
With the result of (\ref{UmRech}) we obtain
\begin{align}
&  \big (\big (\mathcal{W}_{L}^{-1}\otimes\mathcal{W}_{L}^{-1}\big )\circ
\Delta_{L/\bar{L}}\big )\big (\mathcal{W}((x^{1})^{n_{1}}(x^{2})^{n_{2}%
})\big )\nonumber\\
&  =\sum_{i_{1}=0}^{n_{1}}\sum_{i_{2}=0}^{n_{2}}q^{-(n_{1}-\mathcal{\,}%
i_{1})i_{2}}%
\genfrac{[}{]}{0pt}{}{n_{1}}{i_{1}}%
_{q^{-2}}%
\genfrac{[}{]}{0pt}{}{n_{2}}{i_{2}}%
_{q^{-2}}(x^{1})^{i_{1}}(x^{2})^{i_{2}}\otimes(x^{1})^{n_{1}-\,i_{1}}%
(x^{2})^{n_{2}-i_{2}}\nonumber\\
&  \qquad\qquad\times(x^{1})^{i_{1}}(x^{2})^{i_{2}}\otimes(x^{1}%
)^{n_{1}-\,i_{1}}(x^{2})^{n_{2}-i_{2}}\nonumber\\
&  =\sum_{i_{1},i_{2}=0}^{\infty}q^{i_{1}i_{2}}\frac{(x^{1})^{i_{1}}%
(x^{2})^{i_{2}}}{[[i_{1}]]_{q^{-2}}![[i_{2}]]_{q^{-2}}!}\otimes(D_{q^{-2}}%
^{1})^{i_{1}}(D_{q^{-2}}^{2})^{i_{2}}(q^{-i_{2}}x^{1})^{n_{1}}(x^{2})^{n_{2}}.
\label{ComBrai2dim}%
\end{align}
Let us mention that for the second equality we rewrote the q-binomial
coefficients by making use of%
\begin{equation}
(D_{q^{a}}^{\alpha})^{i}(x^{1})^{n_{1}}(x^{2})^{n_{2}}=%
\begin{cases}
\lbrack\lbrack i]]_{q^{a}}!%
\genfrac{[}{]}{0pt}{}{n_{\alpha}}{i}%
_{q^{a}}(x^{\alpha})^{-i}(x^{1})^{n_{1}}(x^{2})^{n_{2}}, & \text{if }0\leq
i<n_{\alpha},\\
0, & \text{if }i>n_{\alpha}\text{.}%
\end{cases}
\label{OpRep2}%
\end{equation}
For the definition of q-numbers, q-factorials, and Jackson derivatives see
Appendix \ref{AppNot}. Realizing that Jackson derivatives are linear operators
it is not very difficult to convince oneself that the operator expression in
Eq. (\ref{ComBrai2dim}) also holds for arbitrary commutative functions (at
least those which can be expanded in an absolutely convergent power series).
Thus, we end up with the result%
\begin{equation}
f(x^{i}\oplus_{L}y^{j})=\sum_{i_{1},i_{2}=0}^{\infty}\frac{(x^{1})^{i_{1}%
}(x^{2})^{i_{2}}}{[[i_{1}]]_{q^{-2}}![[i_{2}]]_{q^{-2}}!}\,\big ((D_{q^{-2}%
}^{1})^{i_{1}}(D_{q^{-2}}^{2})^{i_{2}}f\big )(q^{-i_{2}}y^{1}),
\label{CoForm2dim}%
\end{equation}
which can be viewed as q-deformed Taylor rule in two dimensions.

Now, we will deal on with the operations in (\ref{DefAntiN}). In complete
analogy to our previous considerations we start from the antipode for quantum
space generators. Inserting the expressions for the entries of the L-matrix
into (\ref{SExplN}) we get%
\begin{align}
S_{L}(X^{1})  &  =-\Lambda^{-3/4}\tau^{1/4}X^{1},\nonumber\\
S_{L}(X^{2})  &  =-\Lambda^{-3/4}\tau^{1/4}X^{2}-\lambda\Lambda^{-3/4}%
\tau^{1/4}T^{+}X^{1}.
\end{align}

Again, we try to deduce a formula for the antipode of a normally ordered
monomial. This way, we proceed in the following fashion:
\begin{align}
&  S_{L}((X^{1})^{n_{1}}(X^{2})^{n_{2}})=S_{L}((X^{2})^{n_{2}})S_{L}%
((X^{1})^{n_{1}})\nonumber\\
&  =(S_{L}(X^{2}))^{n_{2}}(S_{L}(X^{1}))^{n_{1}}\nonumber\\
&  =(-1)^{n_{1}+\,n_{2}}(S(\mathcal{L}_{x})_{\alpha}^{2}X^{\alpha})^{n_{2}%
}(S(\mathcal{L}_{x})_{\beta}^{1}X^{\beta})^{n_{1}}. \label{AntUm2dimN}%
\end{align}
Notice that for the first and second equality we used the antihomomorphism
property of the antipode, i.e.
\begin{equation}
S_{L}(a\cdot b)=S_{L}(b)\cdot S_{L}(a).
\end{equation}
From (\ref{DefAntiN}) we see that we have to apply $\mathcal{W}_{R}^{-1}$ to
(\ref{AntUm2dimN}). However, the definition in (\ref{WRh-1}) tells us that we
have to rewrite the last expression of (\ref{AntUm2dimN}) in such a way that
all coordinates stand to the right of all symmetry generators. Using the
commutation relations in (\ref{RelTKoKanfN})-(\ref{RelTKoKendNN}) a
straightforward calculation leads to
\begin{align}
&  \mathcal{W}_{R}^{-1}\big (S_{L}((X^{1})^{n_{1}}(X^{2})^{n_{2}%
})\big )\nonumber\\
&  =(-1)^{n_{1}+\,n_{2}}q^{-n_{1}(n_{1}-1)-n_{2}(n_{2}-1)-n_{1}n_{2}%
}\,\mathcal{W}^{-1}((X^{2})^{n_{2}}(X^{1})^{n_{1}}). \label{AntBraHopN}%
\end{align}
It is not very difficult to read off from (\ref{AntBraHopN}) the operator
expression for an arbitrary commutative function:
\begin{equation}
\hat{U}(f(\ominus_{L}\,x^{i}))=q^{-(\hat{n}_{1})^{2}-(\hat{n}_{2})^{2}-\hat
{n}_{1}\hat{n}_{2}}\,f(-qx^{1},-qx^{2}), \label{AntAll2dim}%
\end{equation}
where the definition of $\hat{n}_{\alpha}$ is given in Appendix \ref{AppNot}.
Notice that the last expression in (\ref{AntBraHopN}) depends on monomials of
reversed ordering. For this reason we introduced the invertible operator
\begin{equation}
\hat{U}(f)\equiv q^{\hat{n}_{1}\hat{n}_{2}}f,
\end{equation}
which transforms a function of ordering $X^{1}X^{2}$ to that representing the
same quantum space element but now for reversed ordering $X^{2}X^{1}$.

It should be clear that we can repeat the same steps as before for the
conjugate Hopf structure \cite{Maj94star, Maj95star}. On quantum space
coordinates the conjugate Hopf structure takes the form%
\begin{align}
\Delta_{\bar{L}}(X^{1})=  &  X^{1}\otimes1+\Lambda^{-3/4}\tau^{1/4}\otimes
X^{1}\nonumber\\
&  +q^{-1}\lambda\Lambda^{-3/4}\tau^{-1/4}T^{-}\otimes X^{2},\nonumber\\
\Delta_{\bar{L}}(X^{2})=  &  X^{2}\otimes1+\Lambda^{-3/4}\tau^{-1/4}\otimes
X^{2},\\[0.16in]
S_{\bar{L}}(X^{1})=  &  -\Lambda^{3/4}\tau^{-1/4}X^{1}+q^{-2}\lambda
\Lambda^{3/4}\tau^{-1/4}T^{-}X^{2},\nonumber\\
S_{\bar{L}}(X^{2})=  &  -\Lambda^{3/4}\tau^{1/4}X^{2}.
\end{align}
From these relations we obtain the formulae
\begin{align}
\Delta_{\bar{L}}((X^{2})^{n_{2}}(X^{1})^{n_{1}})=  &  \sum_{i_{1}=0}^{n_{1}%
}\sum_{i_{2}=0}^{n_{2}}q^{(n_{2}-i_{2})i_{1}}%
\genfrac{[}{]}{0pt}{}{n_{1}}{i_{1}}%
_{q^{2}}%
\genfrac{[}{]}{0pt}{}{n_{2}}{i_{2}}%
_{q^{2}}\nonumber\\
&  \qquad\times(X_{l}^{2})^{i_{2}}(X_{l}^{1})^{i_{1}}(X_{r}^{2})^{n_{2}-i_{2}%
}(X_{r}^{1})^{n_{1}-\,i_{1}},
\end{align}
and
\begin{align}
&  \widetilde{\mathcal{W}}_{R}^{-1}\big (S_{\bar{L}}((X^{2})^{n_{2}}%
(X^{1})^{n_{1}})\big )\nonumber\\
&  =(-1)^{n_{1}+\,n_{2}}q^{n_{1}(n_{1}-1)+n_{2}(n_{2}-1)+n_{1}n_{2}%
}\,\widetilde{\mathcal{W}}^{-1}((X^{1})^{n_{1}}(X^{2})^{n_{2}}),
\end{align}
where%
\begin{gather}
\widetilde{\mathcal{W}}:\mathcal{A}\longrightarrow\mathcal{A}_{q},\nonumber\\
\widetilde{\mathcal{W}}((x^{1})^{i_{1}}(x^{2})^{i_{2}})=(X^{2})^{i_{2}}%
(X^{1})^{i_{1}},
\end{gather}
and%
\begin{gather}
\widetilde{\mathcal{W}}_{R}^{-1}:\mathcal{U}_{q}(su_{2})\ltimes\mathcal{A}%
_{q}\longrightarrow\mathcal{A},\nonumber\\
\widetilde{\mathcal{W}}_{R}^{-1}(h\otimes(X^{2})^{i_{2}}(X^{1})^{i_{1}%
})=\varepsilon(h)\,(x^{1})^{i_{1}}(x^{2})^{i_{2}}.
\end{gather}
The operator expressions for commutative functions now become%
\begin{gather}
f(x^{i}\,\widetilde{\oplus}_{\bar{L}}\,y^{j})=\sum_{i_{1},i_{2}=0}^{\infty
}\frac{(x^{2})^{i_{2}}(x^{1})^{i_{1}}}{[[i_{1}]]_{q^{2}}![[i_{2}]]_{q^{2}}%
!}\,\big ((D_{q^{2}}^{1})^{i_{1}}(D_{q^{2}}^{2})^{i_{2}}f\big )(q^{i_{1}}%
y^{2}),\label{CoKonAll2dim}\\[0.16in]
\hat{U}^{-1}(f(\widetilde{\ominus}_{\bar{L}}\,x^{i}))=q^{(\hat{n}_{1}%
)^{2}+(\hat{n}_{2})^{2}+\hat{n}_{1}\hat{n}_{2}}\,f(-q^{-1}x^{2},-q^{-1}x^{1}),
\label{AntKonAll2dim}%
\end{gather}
where the tilde shall remind us of the fact that the above formulae refer to
reversed normal ordering.

From what we have done so far, we can read off an example for a so-called
crossing symmetry. Comparing the expressions in (\ref{CoForm2dim}) and
(\ref{AntAll2dim}) with those in\ (\ref{CoKonAll2dim}) and
(\ref{AntKonAll2dim}), respectively, shows us the correspondence
\begin{align}
f(x^{i}\oplus_{L}y^{j})  &  \overset{{%
\genfrac{}{}{0pt}{}{\alpha}{q}%
}{%
\genfrac{}{}{0pt}{}{\rightarrow}{\rightarrow}%
}{%
\genfrac{}{}{0pt}{}{\bar{\alpha}}{1/q}%
}}{\longleftrightarrow}f(x^{i}\,\widetilde{\oplus}_{\bar{L}}\,y^{j}%
),\label{TranReg2dim}\\
\hat{U}(f(\ominus_{L}\,x^{i}))  &  \overset{{%
\genfrac{}{}{0pt}{}{\alpha}{q}%
}{%
\genfrac{}{}{0pt}{}{\rightarrow}{\rightarrow}%
}{%
\genfrac{}{}{0pt}{}{\bar{\alpha}}{1/q}%
}}{\longleftrightarrow}\hat{U}^{-1}(f(\widetilde{\ominus}_{\bar{L}}\,x^{i})),
\label{TranReg2dimN}%
\end{align}
where the symbol $\overset{{%
\genfrac{}{}{0pt}{}{\alpha}{q}%
}{%
\genfrac{}{}{0pt}{}{\rightarrow}{\rightarrow}%
}{%
\genfrac{}{}{0pt}{}{\bar{\alpha}}{1/q}%
}}{\longleftrightarrow}$ indicates a transition via the substitutions
\begin{equation}
D_{q^{a}}^{\alpha}\leftrightarrow D_{q^{-a}}^{\bar{\alpha}},\quad\hat
{n}_{\alpha}\leftrightarrow\hat{n}_{\bar{\alpha}},\quad x^{\alpha
}\leftrightarrow x^{\bar{\alpha}},\quad q\leftrightarrow q^{-1},
\end{equation}
with $\alpha=1,2$ and $\bar{\alpha}\equiv3-\alpha.$ It should be emphasized
that in (\ref{TranReg2dim}) and (\ref{TranReg2dimN}) the expressions being
transformed into each other refer to different normal orderings.

Next, we come to translations related to the opposite Hopf structure [cf.
(\ref{RightHopf})]. In this case we have to modify our reasonings slightly.
For the opposite Hopf structure left and right coordinates are now given by
\begin{equation}
X_{l}^{\alpha}\equiv X^{\beta}\otimes(\mathcal{L}_{x})_{\beta}^{\alpha}%
,\quad\text{and\quad}X_{r}^{\alpha}\equiv1\otimes X^{\alpha}.
\end{equation}
Their commutation relations then become
\begin{align}
X_{r}^{1}X_{l}^{1}  &  =q^{2}X_{l}^{1}X_{r}^{1},\nonumber\\
X_{r}^{2}X_{l}^{1}  &  =qX_{l}^{1}X_{r}^{2},\nonumber\\
X_{r}^{2}X_{l}^{2}  &  =q^{2}X_{l}^{2}X_{r}^{2}.
\end{align}
With reasonings similar to those applied to $\Delta_{L}$ we are able to derive
the formula%
\begin{align}
\Delta_{R}((X^{2})^{n_{2}}(X^{1})^{n_{1}})  &  =\sum_{i_{1}=0}^{n_{1}}%
\sum_{i_{2}=0}^{n_{2}}q^{(n_{1}-\,i_{1})i_{2}}%
\genfrac{[}{]}{0pt}{}{n_{1}}{i_{1}}%
_{q^{2}}%
\genfrac{[}{]}{0pt}{}{n_{2}}{i_{2}}%
_{q^{2}}\nonumber\\
&  \qquad\times(X_{l}^{2})^{n_{2}-i_{2}}(X_{l}^{1})^{n_{1}-\,i_{1}}(X_{r}%
^{2})^{i_{2}}(X_{r}^{1})^{i_{1}},
\end{align}
which, in turn, leads to%
\begin{equation}
f(y^{i}\,\widetilde{\oplus}_{R}\,x^{j})=\sum_{i_{1},i_{2}=0}^{\infty}%
\frac{(y^{2})^{i_{2}}(y^{1})^{i_{1}}}{[[i_{1}]]_{q^{2}}![[i_{2}]]_{q^{2}}%
!}\,\big ((D_{q^{2}}^{1})^{i_{1}}(D_{q^{2}}^{2})^{i_{2}}f\big )(q^{i_{2}}%
x^{1}).
\end{equation}

If we start with $\Delta_{\bar{R}}$ and repeat the same steps as before we
will arrive at%
\begin{align}
\Delta_{\bar{R}}((X^{1})^{n_{1}}(X^{2})^{n_{2}})=  &  \sum_{i_{1}=0}^{n_{1}%
}\sum_{i_{2}=0}^{n_{2}}q^{-(n_{2}-i_{2})i_{1}}%
\genfrac{[}{]}{0pt}{}{n_{1}}{i_{1}}%
_{q^{-2}}%
\genfrac{[}{]}{0pt}{}{n_{2}}{i_{2}}%
_{q^{-2}}\nonumber\\
&  \qquad\times(X_{l}^{1})^{n_{1}-\,i_{1}}(X_{l}^{2})^{n_{2}-i_{2}}(X_{r}%
^{1})^{i_{1}}(X_{r}^{2})^{i_{2}},
\end{align}
and
\begin{equation}
f(y^{i}\oplus_{\bar{R}}x^{j})=\sum_{i_{1},i_{2}=0}^{\infty}\frac
{(y^{1})^{i_{1}}(y^{2})^{i_{2}}}{[[i_{1}]]_{q^{-2}}![[i_{2}]]_{q^{-2}}%
!}\,\big ((D_{q^{-2}}^{1})^{i_{1}}(D_{q^{-2}}^{2})^{i_{2}}f\big )(q^{-i_{1}%
}x^{2}).
\end{equation}
With the above results we are now in a position to verify the following
crossing symmetries:
\begin{align}
f(x^{i}\,\widetilde{\oplus}_{R}\,y^{j})  &  \overset{\alpha\leftrightarrow
\bar{\alpha}}{\longleftrightarrow}f(y^{i}\,\widetilde{\oplus}_{\bar{L}}%
\,x^{j}),\nonumber\\
f(x^{i}\oplus_{\bar{R}}y^{j})  &  \overset{\alpha\leftrightarrow\bar{\alpha}%
}{\longleftrightarrow}f(y^{i}\oplus_{L}x^{j}), \label{TransCo2dim}%
\end{align}
where $\overset{\alpha\leftrightarrow\bar{\alpha}}{\longleftrightarrow}$
stands for a transition by means of the substitutions
\begin{equation}
D_{q^{a}}^{\alpha}\leftrightarrow D_{q^{a}}^{\bar{\alpha}},\quad x^{\alpha
}\leftrightarrow x^{\bar{\alpha}},\quad\hat{n}_{\alpha}\leftrightarrow\hat
{n}_{\bar{\alpha}},\quad x\leftrightarrow y.
\end{equation}

Last but not least, we consider the operations that arise from the antipodes
to the opposite Hopf structures. From (\ref{S-1ExplN}) together with the
expressions for the entries of the L-matrices we find%
\begin{align}
S_{L}^{-1}(X^{1})  &  =-X^{1}\tau^{1/4}\Lambda^{-3/4},\nonumber\\
S_{L}^{-1}(X^{2})  &  =-X^{2}\tau^{-1/4}\Lambda^{-3/4}-q\lambda X^{1}T^{+}%
\tau^{-1/4}\Lambda^{-3/4},
\end{align}
and
\begin{align}
S_{\bar{L}}^{-1}(X^{1})  &  =-X^{1}\tau^{1/4}\Lambda^{3/4}-q^{-3}\lambda
X^{2}T^{-}\tau^{1/4}\Lambda^{3/4},\nonumber\\
S_{\bar{L}}^{-1}(X^{2})  &  =-X^{2}\tau^{1/4}\Lambda^{3/4}.
\end{align}
The inverse antipodes of a normally ordered monomial can in general be written
as
\begin{equation}
S_{R}((X^{2})^{n_{2}}(X^{1})^{n_{1}})=(-1)^{n_{1}+\,n_{2}}(X^{\alpha}%
S^{-1}(\mathcal{L}_{x})_{\alpha}^{1})^{n_{1}}(X^{\beta}S^{-1}(\mathcal{L}%
_{x})_{\beta}^{2})^{n_{2}},
\end{equation}
and%
\begin{equation}
S_{\bar{R}}((X^{1})^{n_{1}}(X^{2})^{n_{2}})=(-1)^{n_{1}+\,n_{2}}(X^{\alpha
}S^{-1}(\mathcal{\bar{L}}_{x})_{\alpha}^{1})^{n_{1}}(X^{\beta}S^{-1}%
(\mathcal{\bar{L}}_{x})_{\beta}^{2})^{n_{2}}.
\end{equation}
In the above expressions we have to switch the symmetry generators to the far
left of all coordinates. We are then ready to apply the mappings
$\mathcal{W}_{L}^{-1}$ and $\widetilde{\mathcal{W}}_{L}^{-1}.$ Proceeding this
way, we get
\begin{align}
&  \mathcal{W}_{L}^{-1}\big (S_{R}((X^{2})^{n_{2}}(X^{1})^{n_{1}%
})\big )\nonumber\\
&  =(-1)^{n_{1}+\,n_{2}}q^{n_{1}(n_{1}-1)+n_{2}(n_{2}-1)+n_{1}n_{2}%
}\,\mathcal{W}^{-1}((X^{1})^{n_{1}}(X^{2})^{n_{2}}),
\end{align}
and%
\begin{align}
&  \widetilde{\mathcal{W}}_{L}^{-1}\big (S_{\bar{R}}((X^{1})^{n_{1}}%
(X^{2})^{n_{2}})\big )\nonumber\\
&  =(-1)^{n_{1}+\,n_{2}}q^{-n_{1}(n_{1}-1)-n_{2}(n_{2}-1)-n_{1}n_{2}%
}\,\mathcal{W}^{-1}((X^{2})^{n_{1}}(X^{1})^{n_{1}}).
\end{align}
From the above results it should be obvious that we have%
\begin{equation}
\hat{U}^{-1}(f(\widetilde{\ominus}_{R}\,x^{i}))=q^{(\hat{n}_{1})^{2}+(\hat
{n}_{2})^{2}+\,\hat{n}_{1}\hat{n}_{2}}\,f(-q^{-1}x^{2},-q^{-1}x^{1}),
\end{equation}
and%
\begin{equation}
\hat{U}(f(\ominus_{\bar{R}}\,x^{i}))=q^{-(\hat{n}_{1})^{2}-(\hat{n}_{2}%
)^{2}-\,\hat{n}_{1}\hat{n}_{2}}\,f(-qx^{1},-qx^{2}).
\end{equation}
In complete analogy to (\ref{TransCo2dim}) it holds
\begin{align}
\hat{U}^{-1}(f(\widetilde{\ominus}_{R}\,y^{i}))  &  \overset{\alpha
\leftrightarrow\bar{\alpha}}{\longleftrightarrow}\hat{U}^{-1}(f(\widetilde
{\ominus}_{\bar{L}}\,y^{i})),\nonumber\\
\hat{U}(f(\ominus_{\bar{R}}\,y^{i}))  &  \overset{\alpha\leftrightarrow
\bar{\alpha}}{\longleftrightarrow}\hat{U}(f(\ominus_{L}\,y^{i})),
\end{align}
as can be verified by direct inspection.

\section{Three-dimensional q-deformed Euclidean space\label{Kap2}}

The three-dimensional Euclidean space can be treated along the same line of
arguments as the two-dimensional one. To begin, we consider the conjugate
coproduct on the quantum space coordinates $X^{A},$ $A\in\{+,3,-\}$:%
\begin{align}
\Delta_{\bar{L}}(X^{+})=\,  &  X^{+}\otimes1+\Lambda^{1/2}\tau^{-1/2}\otimes
X^{+},\nonumber\\
\Delta_{\bar{L}}(X^{3})=\,  &  X^{3}\otimes1+\Lambda^{1/2}\otimes
X^{3}+\lambda\lambda_{+}\Lambda^{1/2}L^{-}\otimes X^{+},\nonumber\\
\Delta_{\bar{L}}(X^{-})=\,  &  X^{-}\otimes1+\Lambda^{1/2}\tau^{-1/2}\otimes
X^{-}+q^{-1}\lambda\lambda_{+}\Lambda^{1/2}\tau^{1/2}L^{-}\otimes
X^{3}\nonumber\\
\,  &  +\,q^{-2}\lambda^{2}\lambda_{+}\Lambda^{1/2}\tau^{1/2}(L^{-}%
)^{2}\otimes X^{+}, \label{CoproConN}%
\end{align}
where\ $L^{-}$ and $\tau$ denote generators of $U_{q}(su_{2})$ and $\Lambda$
is a scaling operator. Furthermore, we introduced in (\ref{CoproConN}) the new
parameter $\lambda_{+}\equiv q+q^{-1}.$

Right and left coordinates are defined in complete analogy to the case of the
two-dimensional quantum plane. Thus, from the above relations we can read off
their explicit form. Making use of the commutation relations \cite{LWW97}
\begin{align}
L^{+}X^{+}  &  =X^{+}L^{+},\nonumber\\
L^{+}X^{3}  &  =X^{3}L^{+}-qX^{+}\tau^{-\frac{1}{2}},\nonumber\\
L^{+}X^{-}  &  =X^{-}L^{+}-X^{3}\tau^{-\frac{1}{2}}, \label{VerSu2KoorAnfN}%
\\[0.16in]
L^{-}X^{+}  &  =X^{+}L^{-}+X^{3}\tau^{-\frac{1}{2}},\nonumber\\
L^{-}X^{3}  &  =X^{3}L^{-}+q^{-1}X^{-}\tau^{-\frac{1}{2}},\nonumber\\
L^{-}X^{-}  &  =X^{-}L^{-},\\[0.16in]
\tau^{-\frac{1}{2}}X^{\pm}  &  =q^{\pm2}X^{\pm}\tau^{-\frac{1}{2}},\nonumber\\
\tau^{-\frac{1}{2}}X^{3}  &  =X^{3}\tau^{-\frac{1}{2}},
\label{VertSu2KoorEndN}\\[0.16in]
\Lambda X^{A}  &  =q^{4}X^{A}\Lambda,
\end{align}
we can verify\ that right and left coordinates satisfy
\begin{align}
X_{r}^{+}X_{l}^{+}  &  =q^{4}X_{l}^{+}X_{r}^{+},\nonumber\\
X_{r}^{+}X_{l}^{3}  &  =q^{2}X_{l}^{3}X_{r}^{+},\nonumber\\
X_{r}^{+}X_{l}^{-}  &  =X_{l}^{-}X_{r}^{+},\nonumber\\
X_{r}^{3}X_{l}^{3}  &  =q^{2}X_{l}^{3}X_{r}^{3}+q\lambda\lambda_{+}X_{l}%
^{-}X_{r}^{+},\nonumber\\
X_{r}^{3}X_{l}^{-}  &  =q^{2}X_{l}^{-}X_{r}^{3},\nonumber\\
X_{r}^{-}X_{l}^{-}  &  =q^{4}X_{l}^{-}X_{r}^{-}. \label{ExpVerLinRechN}%
\end{align}
One immediate consequence of these relations is that
\begin{align}
\Delta_{\bar{L}}(X^{+})^{n_{+}}  &  =(X_{l}^{+}+X_{r}^{+})^{n_{+}}=\sum
_{k=0}^{n_{+}}%
\genfrac{[}{]}{0pt}{}{n_{+}}{k}%
_{q^{4}}(X_{l}^{+})^{k}(X_{r}^{+})^{n_{+}-\,k},\label{CoXPl3dim}\\[0.16in]
\Delta_{\bar{L}}(X^{-})^{n_{-}}  &  =(X_{l}^{-}+X_{r}^{-})^{n_{-}}=\sum
_{k=0}^{n_{-}}%
\genfrac{[}{]}{0pt}{}{n_{-}}{k}%
_{q^{4}}(X_{l}^{-})^{k}(X_{r}^{-})^{n_{-}-\,k}. \label{CoXMi3dim}%
\end{align}

To derive an expression for the coproduct on powers of the coordinate $X^{3}$
is a little bit more complicated. To this aim, we make as ansatz
\begin{align}
\Delta_{\bar{L}}(X^{3})^{n_{3}}  &  =(X_{l}^{3}+X_{r}^{3})^{n_{3}}\nonumber\\
&  =\sum_{k=0}^{n_{3}}\sum_{i=0}^{\min(k,n_{3}-k)}C_{k,i}^{n_{3}}\cdot
(X_{l}^{3})^{k-i}(X_{l}^{-})^{i}(X_{r}^{+})^{i}(X_{r}^{3})^{n_{3}-k-i}.
\label{CoX33dimN}%
\end{align}
By exploiting (\ref{ExpVerLinRechN}) we find for the unknown coefficients the
recursion relation
\begin{align}
C_{k,i}^{n_{3}+1}  &  =q^{2(k+i)}C_{k,i}^{n_{3}}+C_{k-1,i}^{n_{3}}\nonumber\\
&  +\,q\lambda\lambda_{+}q^{2(k-i)}[[k-(i-1)]]_{q^{2}}!\,C_{k,i-1}^{n_{3}%
},\nonumber\\
C_{k,0}^{n_{3}}  &  =%
\genfrac{[}{]}{0pt}{}{n_{3}}{k}%
_{q^{2}}.
\end{align}
As one can prove by inserting, the above recursion relation has the solution
\begin{equation}
C_{k,i}^{n_{3}}=(q\lambda\lambda_{+})^{i}\frac{[[k+i]]_{q^{2}}!}%
{[[2i]]_{q^{2}}!![[k-i]]_{q^{2}}!}%
\genfrac{[}{]}{0pt}{}{n_{3}}{k+i}%
_{q^{2}},
\end{equation}
where
\begin{equation}
\lbrack\lbrack2i]]_{q^{2}}!!\equiv\lbrack\lbrack2i]]_{q^{2}}[[2(i-1)]]_{q^{2}%
}\ldots\lbrack\lbrack2]]_{q^{2}}.
\end{equation}

From what we have done so far, it should be obvious that
\begin{align}
&  \Delta_{\bar{L}}((X^{+})^{n_{+}}(X^{3})^{n_{3}}(X^{-})^{n_{-}}%
)=\Delta_{\bar{L}}(X^{+})^{n_{+}}\Delta_{\bar{L}}(X^{3})^{n_{3}}\Delta
_{\bar{L}}(X^{-})^{n_{-}}\nonumber\\
&  =\sum_{k_{+}=0}^{n_{+}}\sum_{k_{3}=0}^{n_{3}}\sum_{k_{-}=0}^{n_{-}}%
\sum_{i=0}^{\min(k_{3},n_{3}-k_{3})}(q\lambda\lambda_{+})^{i}\,\frac
{[[k_{3}+i]]_{q^{2}}!}{[[2i]]_{q^{2}}!![[k_{3}-i]]_{q^{2}}!}\nonumber\\
&  \qquad\times\,q^{2(n_{+}-\,k_{+})(k_{3}-i)+2(n_{3}-i-k_{3})k_{-}}\,%
\genfrac{[}{]}{0pt}{}{n_{+}}{k_{+}}%
_{q^{4}}%
\genfrac{[}{]}{0pt}{}{n_{3}}{k_{3}+i}%
_{q^{2}}%
\genfrac{[}{]}{0pt}{}{n_{-}}{k_{-}}%
_{q^{4}}\nonumber\\
&  \qquad\times\,(X_{l}^{+})^{k_{+}}(X_{l}^{3})^{k_{3}-i}(X_{l}^{-}%
)^{k_{-}+\,i}(X_{r}^{+})^{n_{+}-\,k_{+}+\,i}(X_{r}^{3})^{n_{3}-i-k_{3}}%
(X_{r}^{-})^{n_{-}-\,k_{-}}. \label{CoMon3dim}%
\end{align}
Now, we can follow the same line of arguments as in the previous section.
Applying the mapping
\begin{gather}
\mathcal{W}_{L}^{-1}:\mathcal{U}_{q}(su_{2})\ltimes\mathcal{A}_{q}%
\longrightarrow\mathcal{A},\nonumber\\
\mathcal{W}_{L}^{-1}(h\otimes(X^{+})^{i_{+}}(X^{3})^{i_{3}}(X^{-})^{i_{-}%
})=\varepsilon(h)\,(x^{+})^{i_{+}}(x^{3})^{i_{3}}(x^{-})^{i_{-}},
\end{gather}
to (\ref{CoMon3dim}) and extending the result to commutative functions we
finally obtain
\begin{align}
&  f(x^{A}\oplus_{\bar{L}}y^{B})=\sum_{k_{+}=0}^{n_{+}}\sum_{k_{3}=0}^{n_{3}%
}\sum_{k_{-}=0}^{n_{-}}\sum_{i=0}^{k_{3}}(q\lambda\lambda_{+})^{i}\nonumber\\
&  \qquad\times\frac{(x^{+})^{k_{+}}(x^{3})^{k_{3}-i}(x^{-})^{k_{-}+\,i}%
(y^{+})^{i}}{[[2i]]_{q^{2}}!![[k_{+}]]_{q^{4}}![[k_{3}-i]]_{q^{2}}%
![[k_{-}]]_{q^{4}}!}\nonumber\\
&  \qquad\times\,\big ((D_{q^{4}}^{+})^{k_{+}}(D_{q^{2}}^{3})^{k_{3}%
+i}(D_{q^{4}}^{-})^{k_{-}}f\big )(q^{2(k_{3}-i)}y^{+},q^{2k_{-}}y^{3}).
\label{KomCoprDreiN}%
\end{align}

It is our next goal to find expressions for the operations corresponding to
the antipode $S_{\bar{L}}$. The later is determined by
\begin{align}
S_{\bar{L}}(X^{+})=  &  -\Lambda^{-1/2}\tau^{1/2}X^{+},\nonumber\\
S_{\bar{L}}(X^{3})=  &  -\Lambda^{-1/2}X^{3}+q^{-2}\lambda\lambda_{+}%
\Lambda^{-1/2}\tau^{1/2}L^{-}X^{+},\nonumber\\
S_{\bar{L}}(X^{-})=  &  -\Lambda^{-1/2}\tau^{-1/2}X^{-}+q^{-1}\lambda
\lambda_{+}\Lambda^{-1/2}L^{-}X^{3}\nonumber\\
&  -q^{-4}\lambda^{2}\lambda_{+}\Lambda^{-1/2}\tau^{1/2}(L^{-})^{2}X^{+}.
\end{align}
The Hopf algebra axioms require for the antipode to hold
\begin{equation}
m\circ(S_{\bar{L}}\otimes\text{id})\circ\Delta_{\bar{L}}=\varepsilon,
\end{equation}
where $m$ denotes multiplication on\ the quantum space algebra. Applying these
identities to the coproducts in (\ref{CoXPl3dim}) and (\ref{CoXMi3dim})\ gives
us a system of equations for $S_{\bar{L}}(X^{+})^{n_{+}}$ and $S_{\bar{L}%
}(X^{-})^{n_{-}},$ respectively. Solving these systems by iteration, we get
\begin{align}
S_{\bar{L}}(X^{+})^{n_{+}}  &  =(-1)^{n_{+}}q^{2(n_{+}-1)n_{+}}%
\,m_{\mathcal{H}\otimes A_{q}}((X_{r}^{+})^{n_{+}}),\nonumber\\
S_{\bar{L}}(X^{-})^{n_{-}}  &  =(-1)^{n_{-}}q^{2(n_{-}-1)n_{-}}%
\,m_{\mathcal{H}\otimes A_{q}}((X_{r}^{-})^{n_{-}}),
\end{align}
where%
\begin{equation}
m_{\mathcal{H}\otimes A_{q}}((X_{r}^{\pm})^{n})=(\mathcal{L}_{x})_{A_{1}}%
^{\pm}(\mathcal{L}_{x})_{A_{2}}^{\pm}\ldots(\mathcal{L}_{x})_{A_{n}}^{\pm
}X^{A_{1}}X^{A_{2}}\ldots X^{A_{n}}.
\end{equation}
Taking account of the Hopf algebra axiom%
\begin{equation}
S\circ m=m\circ(S\otimes S)\circ\tau, \label{AntAxiBraiHopf}%
\end{equation}
the method of induction allows us to prove that for the antipode on\ powers of
$X^{3}$ the following expansion holds:
\begin{align}
S_{\bar{L}}(X^{3})^{n_{3}}  &  =\sum_{0\leq2i\leq n_{3}}(-1)^{n_{3}}%
\,(q^{-1}\lambda\lambda_{+})^{i}\,q^{n_{3}(n_{3}-1)-2i(n_{3}-2i)}%
\,[[2i-1]]_{q^{2}}!!\nonumber\\
&  \qquad\qquad\times%
\genfrac{[}{]}{0pt}{}{n_{3}}{2i}%
_{q^{2}}m_{\mathcal{H}\otimes A_{q}}((X^{-})^{i}(X^{3})^{n_{3}-2i}(X^{+}%
)^{i}).
\end{align}
Notice that for the above expression to hold it is required to set
\begin{equation}
\lbrack\lbrack n]]_{q^{2}}!!\equiv1\quad\text{for\quad}n<0.
\end{equation}
Combining everything together and rearranging terms, we should then be able to
obtain
\begin{align}
&  S_{\bar{L}}((X^{+})^{n_{+}}(X^{3})^{n_{3}}(X^{-})^{n_{-}})=S_{\bar{L}%
}((X^{-})^{n_{-}})S_{\bar{L}}((X^{3})^{n_{3}})S_{\bar{L}}((X^{+})^{n_{+}%
})\nonumber\\
\quad &  =(-1)^{n_{+}+\,n_{3}+n_{-}}\,q^{2n_{+}(n_{+}+\,n_{3}-1)+2n_{-}%
(n_{-}+\,n_{3}-1)+n_{3}(n_{3}-1)}\nonumber\\
&  \quad\times\sum_{0\leq2i\leq n_{3}}(q^{-1}\lambda\lambda_{+})^{i}%
\,q^{-2i(n_{3}-2i)}\,[[2i-1]]_{q^{2}}!!\,%
\genfrac{[}{]}{0pt}{}{n_{3}}{2i}%
_{q^{2}}\nonumber\\
&  \quad\qquad\times m_{\mathcal{H}\otimes A_{q}}((X_{r}^{-})^{n_{-}%
+\,i}(X_{r}^{3})^{n_{3}-2i}(X_{r}^{+})^{n_{+}+\,i}).
\end{align}
Repeating the same arguments as in the two-dimensional case, we can arrive at
the expression
\begin{align}
&  \hat{U}(f(\ominus_{\bar{L}}\,x^{A}))=\sum_{i=0}^{\infty}(q^{-1}%
\lambda\lambda_{+})^{i}q^{4i^{2}}\,\frac{(x^{+}x^{-})^{i}}{[[2i]]_{q^{2}}%
!!}\nonumber\\
&  \qquad\times(D_{q^{2}}^{3})^{2i}\,q^{2(\hat{n}_{+}^{2}+\,\hat{n}_{-}%
^{2})+\hat{n}_{3}(2\hat{n}_{+}+\,2\hat{n}_{-}+\,\,\hat{n}_{3}-1)}%
\,f(-x^{+},-q^{-2i}x^{3},-x^{-}), \label{KomAntDreiN}%
\end{align}
where $\hat{U}$ stands for an operator that transforms a function of normal
ordering $X^{+}X^{3}X^{-}$ into another function representing the same element
but now for reversed ordering. The explicit form of this operator was
presented in the work of Ref.\ \cite{BW01}.

We could also have started with the unconjugate Hopf structure, for which we
have%
\begin{align}
\Delta_{L}(X^{+})=\,  &  X^{+}\otimes1+\Lambda^{1/2}\tau^{-1/2}\otimes
X^{+},\nonumber\\
\Delta_{L}(X^{3})=\,  &  X^{3}\otimes1+\Lambda^{1/2}\otimes X^{3}%
+\lambda\lambda_{+}\Lambda^{1/2}L^{-}\otimes X^{+},\nonumber\\
\Delta_{L}(X^{-})=\,  &  X^{-}\otimes1+\Lambda^{1/2}\tau^{-1/2}\otimes
X^{-}+q^{-1}\lambda\lambda_{+}\Lambda^{1/2}\tau^{1/2}L^{-}\otimes
X^{3}\nonumber\\
\,  &  +\,q^{-2}\lambda^{2}\lambda_{+}\Lambda^{1/2}\tau^{1/2}(L^{-}%
)^{2}\otimes X^{+},
\end{align}
and%
\begin{align}
S_{L}(X^{+})=\,  &  -\Lambda^{-1/2}\tau^{1/2}X^{+},\nonumber\\
S_{L}(X^{3})=\,  &  -\Lambda^{-1/2}X^{3}+q^{-2}\lambda\lambda_{+}%
\Lambda^{-1/2}\tau^{1/2}L^{-}X^{+},\nonumber\\
S_{L}(X^{-})=\,  &  -\Lambda^{-1/2}\tau^{-1/2}X^{-}+q^{-1}\lambda\lambda
_{+}\Lambda^{-1/2}L^{-}X^{3}\nonumber\\
\,  &  -\,q^{-4}\lambda^{2}\lambda_{+}\Lambda^{-1/2}\tau^{1/2}(L^{-})^{2}%
X^{+}.
\end{align}
By the same reasonings as above we are led to the expressions
\begin{align}
&  \Delta_{L}((X^{-})^{n_{-}}(X^{3})^{n_{3}}(X^{+})^{n_{+}})=\sum_{k_{+}%
=0}^{n_{+}}\sum_{k_{3}=0}^{n_{3}}\sum_{k_{-}=0}^{n_{-}}\sum_{i=0}^{\min
(k_{3},n_{3}-k_{3})}(-q^{-1}\lambda\lambda_{+})^{i}\nonumber\\
&  \,\qquad\times\,q^{-2(n_{-}-\,k_{-})(k_{3}-i)-2(n_{3}-i-k_{3})k_{+}}%
\,\frac{[[k_{3}+i]]_{q^{-2}}!}{[[2i]]_{q^{-2}}!![[k_{3}-i]]_{q^{-2}}%
!}\nonumber\\
&  \qquad\times%
\genfrac{[}{]}{0pt}{}{n_{+}}{k_{+}}%
_{q^{-4}}%
\genfrac{[}{]}{0pt}{}{n_{3}}{k_{3}+i}%
_{q^{-2}}%
\genfrac{[}{]}{0pt}{}{n_{-}}{k_{-}}%
_{q^{-4}}(X_{l}^{-})^{k_{-}}(X_{l}^{3})^{k_{3}-i}(X_{l}^{+})^{k_{+}%
+\,i}\nonumber\\
&  \qquad\times(X_{r}^{-})^{n_{-}-\,k_{-}+\,i}(X_{r}^{3})^{n_{3}-i-k_{3}%
}(X_{r}^{+})^{n_{+}-\,k_{+}},
\end{align}
and
\begin{align}
&  S_{L}((X^{-})^{n_{-}}(X^{3})^{n_{3}}(X^{+})^{n_{+}})\nonumber\\
&  \quad=(-1)^{n_{+}+\,n_{3}+n_{-}}\,q^{-2n_{-}(n_{-}+\,n_{3}-1)-2n_{+}%
(n_{+}+\,n_{3}-1)-n_{3}(n_{3}-1)}\nonumber\\
&  \quad\times\sum_{0\leq2k\leq n_{3}}(-q\lambda\lambda_{+})^{k}%
\,q^{2k(n_{3}-2k)}\,[[2k-1]]_{q^{-2}}!!\,%
\genfrac{[}{]}{0pt}{}{n_{3}}{2k}%
_{q^{-2}}\nonumber\\
&  \quad\qquad\times(X_{r}^{+})^{n_{+}+\,k}(X_{r}^{3})^{n_{3}-2k}(X_{r}%
^{-})^{n_{-}+\,k}.
\end{align}
The corresponding operations on commutative functions take on the form
\begin{align}
&  f(x^{A}\,\widetilde{\oplus}_{L}\,y^{B})=\sum_{k_{+}=0}^{n_{+}}\sum
_{k_{3}=0}^{n_{3}}\sum_{k_{-}=0}^{n_{-}}\sum_{i=0}^{k_{3}}(-q^{-1}%
\lambda\lambda_{+})^{i}\nonumber\\
&  \qquad\times\frac{(x^{-})^{k_{-}+\,i}(x^{3})^{k_{3}-i}(x^{+})^{k_{+}%
+\,i}(y^{-})^{i}}{[[2i]]_{q^{-2}}!![[k_{+}]]_{q^{-4}}![[k_{3}-i]]_{q^{-2}%
}![[k_{-}]]_{q^{-4}}!}\nonumber\\
&  \qquad\times\big ((D_{q^{-4}}^{-})^{k_{-}}(D_{q^{-2}}^{3})^{k_{3}%
+i}(D_{q^{-4}}^{+})^{k_{+}}f\big )(q^{-2(k_{3}-i)}y^{-},q^{-2k_{+}}y^{3}),
\end{align}
and
\begin{align}
&  \hat{U}^{-1}(f(\widetilde{\ominus}_{L}\,x^{A}))=\sum_{k=0}^{\infty
}(-q\lambda\lambda_{+})^{k}q^{-4k^{2}}\,\frac{(x^{+}x^{-})^{k}}{[[2k]]_{q^{-2}%
}!!}\nonumber\\
&  \qquad\times(D_{q^{-2}}^{3})^{2k}\,q^{-2(\hat{n}_{+}^{2}+\,\hat{n}_{-}%
^{2})-\hat{n}_{3}(2\hat{n}_{+}+\,2\hat{n}_{-}+\,\,\hat{n}_{3}-1)}%
\,f(-x^{-},-q^{2k}x^{3},-x^{+}).
\end{align}
The tilde shall again remind us of the fact that the above operations refer to
reversed normal ordering, i.e. $X^{-}X^{3}X^{+}.$

Comparing the above results with (\ref{KomCoprDreiN}) and (\ref{KomAntDreiN})
shows us the existence of the crossing symmetries
\begin{equation}
f(x^{A}\oplus_{\bar{L}}y^{B})\overset{{%
\genfrac{}{}{0pt}{}{\pm}{q}%
}{%
\genfrac{}{}{0pt}{}{\rightarrow}{\rightarrow}%
}{%
\genfrac{}{}{0pt}{}{\mp}{1/q}%
}}{\longleftrightarrow}f(x^{A}\,\widetilde{\oplus}_{L}\,y^{B}),
\end{equation}
and
\begin{equation}
\hat{U}(f(\ominus_{\bar{L}}\,x^{A}))\overset{{%
\genfrac{}{}{0pt}{}{\pm}{q}%
}{%
\genfrac{}{}{0pt}{}{\rightarrow}{\rightarrow}%
}{%
\genfrac{}{}{0pt}{}{\mp}{1/q}%
}}{\longleftrightarrow}\hat{U}^{-1}(f(\widetilde{\ominus}_{L}\,x^{A})),
\end{equation}
where the symbol $\overset{{%
\genfrac{}{}{0pt}{}{\pm}{q}%
}{%
\genfrac{}{}{0pt}{}{\rightarrow}{\rightarrow}%
}{%
\genfrac{}{}{0pt}{}{\mp}{1/q}%
}}{\longleftrightarrow}$ indicates a transition via the substitutions
\begin{equation}
D_{q^{a}}^{\pm}\leftrightarrow D_{q^{-a}}^{\mp},\quad\hat{n}_{\pm
}\leftrightarrow\hat{n}_{\mp},\quad x^{\pm}\leftrightarrow x^{\mp},\quad
q\leftrightarrow q^{-1}.
\end{equation}

We can also perform our calculations for the opposite Hopf structure. In doing
so, we end up at expressions that can be obtained most easily from our
previous results by applying the crossing symmetries
\begin{align}
f(x^{A}\oplus_{\bar{L}}y^{B})  &  \overset{+\leftrightarrow-}%
{\longleftrightarrow}f(y^{A}\oplus_{R}x^{B}),\nonumber\\
f(x^{A}\,\widetilde{\oplus}_{\bar{R}}\,y^{B})  &  \overset{+\leftrightarrow
-}{\longleftrightarrow}f(y^{A}\,\widetilde{\oplus}_{L}\,x^{B}),
\end{align}
and
\begin{align}
\hat{U}(f(\ominus_{\bar{L}}\,x^{A}))  &  \overset{+\leftrightarrow
-}{\longleftrightarrow}\hat{U}(f(\ominus_{R}\,x^{A})),\nonumber\\
\hat{U}^{-1}(f(\widetilde{\ominus}_{\bar{R}}\,x^{A}))  &  \overset
{+\leftrightarrow-}{\longleftrightarrow}\hat{U}^{-1}(f(\widetilde{\ominus}%
_{L}\,x^{A})),
\end{align}
where symbol $\overset{+\leftrightarrow-}{\longleftrightarrow}$ stands for the
substitutions
\begin{equation}
D_{q^{a}}^{\pm}\leftrightarrow D_{q^{a}}^{\mp},\quad\hat{n}_{\pm
}\leftrightarrow\hat{n}_{\mp}\quad x^{\pm}\leftrightarrow x^{\mp},\quad
x\leftrightarrow y.
\end{equation}

\section{Four-dimensional q-deformed Euclidean space\label{4dimEuc}}

All considerations of the previous sections pertain equally to the q-deformed
Euclidean space with four dimensions. Thus we can restrict ourselves to
stating the results. Again, we start from the coproduct on the quantum space
coordinates $X^{i},$ $i=1,\ldots,4$:
\begin{align}
\Delta_{L}(X^{1})  &  =X^{1}\otimes1+\Lambda^{-1/2}K_{1}^{1/2}K_{2}%
^{1/2}\otimes X^{1},\nonumber\\
\Delta_{L}(X^{2})  &  =X^{2}\otimes1+\Lambda^{-1/2}K_{1}^{-1/2}K_{2}%
^{1/2}\otimes X^{2}\nonumber\\
&  +\,q\lambda\Lambda^{-1/2}K_{1}^{1/2}K_{2}^{1/2}L_{1}^{+}\otimes
X^{1},\nonumber\\
\Delta_{L}(X^{3})  &  =X^{3}\otimes1+\Lambda^{-1/2}K_{1}^{1/2}K_{2}%
^{-1/2}\otimes X^{3}\nonumber\\
&  +\,q\lambda\Lambda^{-1/2}K_{1}^{1/2}K_{2}^{1/2}L_{2}^{+}\otimes
X^{1},\nonumber\\
\Delta_{L}(X^{4})  &  =X^{4}\otimes1+\Lambda^{-1/2}K_{1}^{-1/2}K_{2}%
^{-1/2}\otimes X^{4}\nonumber\\
&  -\,q^{2}\lambda^{2}\Lambda^{-1/2}K_{1}^{1/2}K_{2}^{1/2}L_{1}^{+}L_{2}%
^{+}\otimes X^{1}\nonumber\\
&  -\,q\lambda\Lambda^{-1/2}K_{1}^{-1/2}K_{2}^{1/2}L_{2}^{+}\otimes
X^{2}\nonumber\\
&  -\,q\lambda\Lambda^{-1/2}K_{1}^{1/2}K_{2}^{-1/2}L_{1}^{+}\otimes X^{3},
\label{Ant4dimUnco}%
\end{align}
where $K_{i}$ and $L_{i}^{-},$ $i=1,2,$ are generators of the Hopf algebra
$U_{q}(so_{4}).$ As usual, these coproducts allow us to read off the explicit
form of right and left coordinates. With the help of the commutation relations
between symmetry generators and coordinates (see for example Ref. \cite{BW01})
one verifies the relations
\begin{align}
X_{r}^{i}X_{l}^{i}  &  =q^{-2}X_{l}^{i}X_{r}^{i},\text{\quad}i=1,\ldots
,4,\nonumber\\
X_{r}^{m}X_{l}^{n}  &  =q^{-1}X_{l}^{n}X_{r}^{m},\quad(m,n)\in
\{(1,2),(1,3),(2,4),(3,4)\},\nonumber\\
X_{r}^{2}X_{l}^{3}  &  =X_{l}^{3}X_{r}^{2}+\lambda X_{l}^{4}X_{r}%
^{1},\nonumber\\
X_{r}^{1}X_{l}^{4}  &  =X_{l}^{4}X_{r}^{1}. \label{BraRel4dimN}%
\end{align}
Using the same reasonings already applied to the two- and three-dimensional
Euclidean space, we can then derive the following q-binomial rule from the
above relations:%
\begin{align}
&  \Delta_{L}((X^{1})^{n_{1}}(X^{2})^{n_{2}}(X^{3})^{n_{3}}(X^{4})^{n_{4}%
})\nonumber\\
&  =\sum_{k_{1}=0}^{n_{1}}\sum_{k_{2}=0}^{n_{2}}\sum_{k_{3}=0}^{n_{3}}%
\sum_{k_{4}=0}^{n_{4}}\sum_{i=0}^{\min(n_{2}-k_{2},k_{3})}\lambda
^{i}\,q^{-(k_{2}+k_{3}-i)(n_{1}-\,k_{1})-k_{4}(n_{2}+n_{3}-k_{2}-k_{3}%
-i)}\nonumber\\
&  \qquad\times\,[[i]]_{q^{-2}}!\,%
\genfrac{[}{]}{0pt}{}{k_{2}+i}{i}%
_{q^{-2}}%
\genfrac{[}{]}{0pt}{}{k_{3}}{i}%
_{q^{-2}}%
\genfrac{[}{]}{0pt}{}{n_{1}}{k_{1}}%
_{q^{-2}}%
\genfrac{[}{]}{0pt}{}{n_{2}}{k_{2}+i}%
_{q^{-2}}%
\genfrac{[}{]}{0pt}{}{n_{3}}{k_{3}}%
_{q^{-2}}%
\genfrac{[}{]}{0pt}{}{n_{4}}{k_{4}}%
_{q^{-2}}\nonumber\\
&  \qquad\times\,(X_{l}^{1})^{k_{1}}(X_{l}^{2})^{k_{2}}(X_{l}^{3})^{k_{3}%
-i}(X_{l}^{4})^{k_{4}+i}\nonumber\\
&  \qquad\times\,(X_{r}^{1})^{n_{1}-\,k_{1}+\,i}(X_{r}^{2})^{n_{2}-k_{2}%
-i}(X_{r}^{3})^{n_{3}-k_{3}}(X_{r}^{4})^{n_{4}-k_{4}}. \label{qBin4dim}%
\end{align}
If we define
\begin{gather}
\mathcal{W}_{L}^{-1}:\mathcal{U}_{q}(so_{4})\ltimes\mathcal{A}_{q}%
\longrightarrow\mathcal{A},\nonumber\\[0.1in]
\mathcal{W}_{L}^{-1}(h\otimes(X^{1})^{i_{1}}(X^{2})^{i_{2}}(X^{3})^{i_{3}%
}(X^{4})^{i_{4}})\nonumber\\
=\varepsilon(h)\,(x^{1})^{i_{1}}(x^{2})^{i_{2}}(x^{3})^{i_{3}}(x^{4})^{i_{4}},
\end{gather}
the result in (\ref{qBin4dim}) implies%
\begin{align}
&  f(x^{i}\oplus_{L}y^{j})\nonumber\\
&  =\sum_{k_{1}=0}^{\infty}\sum_{k_{2}=0}^{\infty}\sum_{k_{3}=0}^{\infty}%
\sum_{k_{4}=0}^{\infty}\sum_{i=0}^{k_{3}}\lambda^{i}%
\genfrac{[}{]}{0pt}{}{k_{3}}{i}%
_{q^{-2}}\frac{(x^{1})^{k_{1}}(x^{2})^{k_{2}}(x^{3})^{k_{3}-i}(x^{4}%
)^{k_{4}+i}(y^{1})^{i}}{[[k_{1}]]_{q^{-2}}![[k_{2}]]_{q^{-2}}![[k_{3}%
]]_{q^{-2}}![[k_{4}]]_{q^{-2}}!}\nonumber\\
&  \quad\times\big((D_{q^{-2}}^{1})^{k_{1}}(D_{q^{-2}}^{2})^{k_{2}%
+i}(D_{q^{-2}}^{3})^{k_{3}}(D_{q^{-2}}^{4})^{k_{4}}f\big )(q^{-(k_{2}%
+k_{3}+i)}y^{1},q^{-k_{4}}y^{2},q^{-k_{4}}y^{3}).
\end{align}

Next we deal on with the antipode corresponding to (\ref{Ant4dimUnco}). On the
coordinates $X^{i},$ $i=1,\ldots,4,$ it becomes%
\begin{align}
S_{L}(X^{1})=  &  -\Lambda^{1/2}K_{1}^{-1/2}K_{2}^{-1/2}X^{1},\nonumber\\
S_{L}(X^{2})=  &  -\Lambda^{1/2}K_{1}^{1/2}K_{2}^{-1/2}(X^{2}-q^{2}\lambda
L_{1}^{+}X^{1}),\nonumber\\
S_{L}(X^{3})=  &  -\Lambda^{1/2}K_{1}^{-1/2}K_{2}^{1/2}(X^{3}-q^{2}\lambda
L_{2}^{+}X^{1}),\nonumber\\
S_{L}(X^{4})=  &  -\Lambda^{1/2}K_{1}^{1/2}K_{2}^{1/2}(X^{4}+q^{2}%
\lambda(L_{1}^{+}X^{3}+L_{2}^{+}X^{2}))\nonumber\\
&  -\,q^{4}\lambda^{2}\Lambda^{1/2}K_{1}^{1/2}K_{2}^{1/2}L_{1}^{+}L_{2}%
^{+}X^{1}.
\end{align}
In complete analogy to the three-dimensional case we find the expression
\begin{align}
&  \hspace{-0.16in}S_{L}((X^{1})^{n_{1}}(X^{2})^{n_{2}}(X^{3})^{n_{3}}%
(X^{4})^{n_{4}})\nonumber\\
=  &  \,(-1)^{n_{1}+\,n_{2}+n_{3}+n_{4}}\,q^{-n_{1}(n_{1}-1)-n_{2}%
(n_{2}-1)-n_{3}(n_{3}-1)-n_{4}(n_{4}-1)}\nonumber\\
&  \,\times\sum_{k=0}^{\min(n_{2},n_{3})}\lambda^{k}\,q^{-\frac{1}%
{2}k(k+1)-k^{2}+(n_{1}+n_{4}+k)(n_{2}+n_{3})}\,[[k]]_{q^{-2}}!\,%
\genfrac{[}{]}{0pt}{}{n_{2}}{k}%
_{q^{-2}}%
\genfrac{[}{]}{0pt}{}{n_{3}}{k}%
_{q^{-2}}\nonumber\\
&  \,\qquad\qquad\quad\times\,m_{\mathcal{H}\otimes A_{q}}((X_{r}^{4}%
)^{n_{4}+k}(X_{r}^{3})^{n_{3}-k}(X_{r}^{2})^{n_{2}-k}(X_{r}^{1})^{n_{1}+k}).
\end{align}
This result corresponds to the operator%
\begin{align}
&  \hspace{-0.16in}\hat{U}(f(\ominus_{L}x^{i}))=\sum_{k=0}^{\infty}\lambda
^{k}q^{-\frac{1}{2}k(k+1)-k^{2}}\,\frac{(x^{1}x^{4})^{k}}{[[k]]_{q^{-2}}%
!}\,(D_{q^{-2}}^{2}D_{q^{-2}}^{3})^{k}\nonumber\\
&  \qquad\times\,q^{-\hat{n}_{1}(\hat{n}_{1}-1)-\hat{n}_{2}(\hat{n}%
_{2}-1)-\hat{n}_{3}(\hat{n}_{3}-1)-\hat{n}_{4}(\hat{n}_{4}-1)-(\hat{n}%
_{1}+\hat{n}_{2})(\hat{n}_{3}+\hat{n}_{4})}\nonumber\\
&  \qquad\times\,f(-x^{1},-q^{k}x^{2},-q^{k}x^{3},-x^{4}).
\end{align}
Again, we introduced an\ operator $\hat{U}$ that transforms functions
referring to ordering $X^{1}X^{2}X^{3}X^{4}$ to those of reversed ordering
$X^{4}X^{3}X^{2}X^{1}$. Its explicit form can be looked up in Ref. \cite{BW01}.

As we know, our considerations also apply to the conjugate Hopf structure,
which on coordinates $X^{i},$ $i=1,\ldots,4,$ explicitly reads%
\begin{align}
\Delta_{\bar{L}}(X^{1})=\, &  X^{1}\otimes1+\Lambda^{1/2}K_{1}^{-1/2}%
K_{2}^{-1/2}\otimes X^{1}\nonumber\\
\, &  -\,q^{-1}\lambda\Lambda^{1/2}K_{1}^{1/2}K_{2}^{-1/2}L_{1}^{-}\otimes
X^{2}\nonumber\\
\, &  -\,q^{-1}\lambda\Lambda^{1/2}K_{1}^{-1/2}K_{2}^{1/2}L_{2}^{-}\otimes
X^{3}\nonumber\\
\, &  -\,q^{-2}\lambda^{2}\Lambda^{1/2}K_{1}^{1/2}K_{2}^{1/2}L_{1}^{-}%
L_{2}^{-}\otimes X^{4},\nonumber\\
\Delta_{\bar{L}}(X^{2})=\, &  X^{2}\otimes1+\Lambda^{1/2}K_{1}^{1/2}%
K_{2}^{-1/2}\otimes X^{2}\nonumber\\
\, &  +\,q^{-1}\lambda\Lambda^{1/2}K_{1}^{1/2}K_{2}^{1/2}L_{2}^{-}\otimes
X^{4},\nonumber\\
\Delta_{\bar{L}}(X^{3})=\, &  X^{3}\otimes1+\Lambda^{1/2}K_{1}^{-1/2}%
K_{2}^{1/2}\otimes X^{3}\nonumber\\
\, &  +\,q^{-1}\lambda\Lambda^{1/2}K_{1}^{1/2}K_{2}^{1/2}L_{1}^{-}\otimes
X^{4},\nonumber\\
\Delta_{\bar{L}}(X^{4})=\, &  X^{4}\otimes1+\Lambda^{1/2}K_{1}^{1/2}%
K_{2}^{1/2}\otimes X^{4},
\end{align}
and%
\begin{align}
S_{\bar{L}}(X^{1})=\, &  -\Lambda^{-1/2}K_{1}^{1/2}K_{2}^{1/2}(X^{1}%
+q^{-2}\lambda(L_{1}^{-}X^{2}+L_{2}^{-}X^{3}))\nonumber\\
\, &  +\,q^{-4}\lambda^{2}\Lambda^{1/2}K_{1}^{1/2}K_{2}^{1/2}L_{1}^{-}%
L_{2}^{-}X^{4},\nonumber\\
S_{\bar{L}}(X^{2})=\, &  -\Lambda^{-1/2}K_{1}^{-1/2}K_{2}^{1/2}(X^{2}%
-q^{-2}\lambda L_{2}^{-}X^{4}),\nonumber\\
S_{\bar{L}}(X^{3})=\, &  -\Lambda^{-1/2}K_{1}^{-1/2}K_{2}^{1/2}(X^{3}%
-q^{-2}\lambda L_{1}^{-}X^{4}),\nonumber\\
S_{\bar{L}}(X^{4})=\, &  -\Lambda^{-1/2}K_{1}^{-1/2}K_{2}^{-1/2}X^{4}.
\end{align}
We can also perform our calculations starting from\ opposite Hopf structures.
Obviously, each of these Hopf structures leads to its own q-translation. The
results for the different Hopf structures are related to each other by the
correspondences
\begin{align}
f(x^{i}\,\widetilde{\oplus}_{\bar{L}}\,y^{j}) &  \overset{{{%
\genfrac{}{}{0pt}{}{i}{q}%
}{%
\genfrac{}{}{0pt}{}{\rightarrow}{\rightarrow}%
}{%
\genfrac{}{}{0pt}{}{i^{\prime}}{1/q}%
}}}{\longleftrightarrow}f(x^{i}\oplus_{L}y^{j}),\\[0.1in]
f(x^{i}\,\widetilde{\oplus}_{\bar{L}}\,y^{j}) &  \overset{i\leftrightarrow
i^{\prime}}{\longleftrightarrow}f(y^{i}\,\widetilde{\oplus}_{R}\,x^{j}%
),\nonumber\\
f(x^{i}\oplus_{\bar{R}}y^{j}) &  \overset{i\leftrightarrow i^{\prime}%
}{\longleftrightarrow}f(y^{i}\oplus_{L}x^{j}),
\end{align}
and
\begin{align}
\hat{U}^{-1}(f(\widetilde{\ominus}_{\bar{L}}\,x^{i})) &  \overset{{{%
\genfrac{}{}{0pt}{}{i}{q}%
}{%
\genfrac{}{}{0pt}{}{\rightarrow}{\rightarrow}%
}{%
\genfrac{}{}{0pt}{}{\overline{i}}{1/q}%
}}}{\longleftrightarrow}\hat{U}(f(\ominus_{L}\,x^{i})),\\[0.1in]
\hat{U}^{-1}(f(\widetilde{\ominus}_{\bar{L}}\,x^{i})) &  \overset
{i\leftrightarrow\overline{i}}{\longleftrightarrow}\hat{U}^{-1}(f(\widetilde
{\ominus}_{R}\,x^{i})),\nonumber\\
\hat{U}(f(\ominus_{\bar{R}}\,x^{i})) &  \overset{i\leftrightarrow\overline{i}%
}{\longleftrightarrow}\hat{U}(f(\ominus_{L}\,x^{i})).
\end{align}
The symbol $\overset{{{%
\genfrac{}{}{0pt}{}{i}{q}%
}{%
\genfrac{}{}{0pt}{}{\rightarrow}{\rightarrow}%
}{%
\genfrac{}{}{0pt}{}{\overline{i}}{1/q}%
}}}{\longleftrightarrow}$ now denotes the substitutions ($\overline{i}%
\equiv5-i)$%
\begin{equation}
D_{q^{a}}^{i}\leftrightarrow D_{q^{-a}}^{\overline{i}},\quad\hat{n}%
_{i}\leftrightarrow\hat{n}_{\overline{i}},\quad x^{i}\leftrightarrow
x^{i},\quad q\leftrightarrow q^{-1},
\end{equation}
and $\overset{i\leftrightarrow\overline{i}}{\longleftrightarrow}$ stands for a
transition via
\begin{equation}
D_{q^{a}}^{i}\leftrightarrow D_{q^{a}}^{\overline{i}},\quad\hat{n}%
_{i}\leftrightarrow\hat{n}_{\overline{i}},\quad x^{i}\leftrightarrow
x^{\overline{i}},\quad x\leftrightarrow y.
\end{equation}

\section{q-Deformed Minkowski space\label{MinSpace}}

In this section we would like to deal with q-deformed Minkowski space, which
from a physical point of view is the most interesting quantum space.
Unfortunately, if we try to\ proceed in the same way as in the previous
sections we get expressions of a\ rather complicated structure. For this
reason we would like to present another method for calculating q-translations.

To this end, let us first recall the abstract definition of\ an exponential
\cite{Maj93-5}. We assume that to each quantum space $\mathcal{A}_{q}$ exists
a dual object $\mathcal{A}_{q}^{\ast}$ with pairings%
\begin{equation}
\left\langle .\,,.\right\rangle :\mathcal{A}_{q}^{\ast}\otimes\mathcal{A}%
_{q}\rightarrow\mathbb{C}\text{,}\quad\text{\quad}\big \langle f^{b}%
,e_{a}\big \rangle=\delta_{a}^{b}, \label{DefParAb1}%
\end{equation}
and%
\begin{equation}
\left\langle .\,,.\right\rangle ^{\prime}:\mathcal{A}_{q}\otimes
\mathcal{A}_{q}^{\ast}\rightarrow\mathbb{C}\text{,}\quad\text{\quad
}\big \langle e_{a},f^{b}\big \rangle^{\prime}=\delta_{a}^{b},
\label{DefParAb2}%
\end{equation}
where $\{e_{a}\}$ is a basis in $\mathcal{A}_{q}$ and $\{f^{b}\}$ a dual basis
in $\mathcal{A}_{q}^{\ast}.$ An exponential is nothing other than a mapping
whose dualisation is given by one of the pairings in (\ref{DefParAb1}) and
(\ref{DefParAb2}). More concretely, we have%
\begin{equation}
\exp:\mathbb{C}\longrightarrow\mathcal{A}_{q}\otimes\mathcal{A}_{q}^{\ast
},\quad\text{\quad}\exp(\lambda)=\lambda\sum_{a}e_{a}\otimes f^{a},
\label{AbsExp1}%
\end{equation}
or%
\begin{equation}
\exp^{\prime}:\mathbb{C}\longrightarrow\mathcal{A}_{q}^{\ast}\otimes
\mathcal{A}_{q},\quad\text{\quad}\exp^{\prime}(\lambda)=\lambda\sum_{a}%
f^{a}\otimes e_{a}. \label{AbsExp2}%
\end{equation}

In Ref. \cite{Maj93-5} it was shown that the algebra of quantum space
coordinates and that of the corresponding partial derivatives are dual to each
other. The dual pairings are given by
\begin{align}
\left\langle .\,,.\right\rangle :\mathcal{A}_{q}^{\ast}\otimes\mathcal{A}%
_{q}\rightarrow\mathbb{C}\quad\text{with\quad}\big \langle f(\partial
^{i}),g(X^{j})\big \rangle &  \equiv\varepsilon(f(\partial^{i})\triangleright
g(X^{j}))\nonumber\\
&  =\varepsilon(f(\partial^{i})\,\bar{\triangleleft}\,g(X^{j}%
)),\label{DefPar1N}%
\end{align}
and%
\begin{align}
\left\langle .\,,.\right\rangle ^{\prime}:\mathcal{A}_{q}\otimes
\mathcal{A}_{q}^{\ast}\rightarrow\mathbb{C}\quad\text{with\quad}\big \langle
f(X^{i}),g(\partial^{j})\big \rangle^{\prime} &  \equiv\varepsilon
(f(X^{i})\triangleleft g(\partial^{j}))\nonumber\\
&  =\varepsilon(f(X^{i})\,\bar{\triangleright}\,g(\partial^{j}%
)).\label{DefPar2N}%
\end{align}
To find the explicit form of the corresponding exponentials it is our task to
determine a basis of the coordinate algebra $\mathcal{A}_{q}$ being dual to a
given one of the derivative algebra $\mathcal{A}_{q}^{\ast}$. Inserting the
elements of the two bases into the defining expressions in\ (\ref{AbsExp1})
and (\ref{AbsExp2}) will then provide us with formulae for exponentials on
quantum spaces. Via the algebra isomorphism $\mathcal{W}$ [cf. Eq.
(\ref{AlgIsoN})] we are then able to introduce q-deformed exponentials living
on tensor products of commutative algebras. More concretely, this is achieved
by the expressions%
\begin{equation}
\exp(x^{i}|\partial^{j})\equiv\sum_{a}\mathcal{W}(e_{a})\otimes\mathcal{W}%
(f^{a}),\label{ExpAll}%
\end{equation}
and%
\begin{equation}
\exp(\partial^{i}|x^{j})\equiv\sum_{a}\mathcal{W}(f^{a})\otimes\mathcal{W}%
(e_{a}).\label{ExpAl2}%
\end{equation}
In Ref. \cite{Wac03} we used this method for calculating q-deformed exponentials.

It should be recorded that the exponentials in (\ref{ExpAll}) and
(\ref{ExpAl2}) can also be viewed as q-analogs of classical plane waves, since
they obey
\begin{align}
\partial^{i}\overset{x}{\triangleright}\exp(x^{k}|\partial^{l})  &
=\exp(x^{k}|\partial^{l})\overset{\partial}{\circledast}\partial
^{i},\nonumber\\
\exp(\partial^{l}|x^{k})\overset{x}{\triangleleft}\partial^{i}  &
=\partial^{i}\overset{\partial}{\circledast}\exp(\partial^{l}|x^{k}),
\label{EigGl1N}%
\end{align}
and%
\begin{align}
x^{i}\overset{x}{\circledast}\exp(x^{k}|\partial^{l})  &  =\exp(x^{k}%
|\partial^{l})\overset{\partial}{\triangleleft}x^{i},\nonumber\\
\exp(\partial^{l}|x^{k})\overset{x}{\circledast}x^{i}  &  =x^{i}%
\overset{\partial}{\triangleright}\exp(\partial^{l}|x^{k}). \label{EigGl2}%
\end{align}
In addition to this they fulfill the normalization condition%
\begin{equation}
\exp(x^{i}|\partial^{j})|_{x^{i}=0}=\exp(x^{i}|\partial^{j})|_{\partial^{j}%
=0}=1. \label{NorExpN}%
\end{equation}
For the details we refer the reader to Refs. \cite{Wac03} and \cite{Maj93-5}.

Important for us is the fact that q-deformed exponentials generate q-deformed
translations in the following way:%
\begin{align}
\exp(x^{i}|\partial^{j})\overset{\partial|y}{\triangleright}g(y^{k})  &
=g(x^{i}\oplus y^{k}),\nonumber\\
g(y^{k})\overset{y|\partial}{\triangleleft}\exp(\partial^{j}|x^{i})  &
=g(y^{k}\oplus x^{i}), \label{q-TayN}%
\end{align}
Notice that in (\ref{q-TayN}) the derivatives being part of the q-deformed
exponential act upon the function $g(y^{k})$. To prove the above identities we
first recall that the action upon partial derivatives can be written as%
\begin{align}
\partial^{i}\triangleright g  &  =\big \langle\partial^{i},\mathcal{W}%
^{-1}(g_{(1)})\big \rangle g_{(2)},\nonumber\\
g\triangleleft\partial^{i}  &  =g_{(2)}\big \langle\mathcal{W}^{-1}%
(g_{(1)}),\partial^{i}\big \rangle^{\prime}, \label{CorActN}%
\end{align}
where
\begin{equation}
g_{(1)}\otimes g_{(2)}\equiv g(x^{i}\oplus y^{k}).
\end{equation}
Now, we can proceed as follows:%
\begin{align}
\exp(x^{i}|\partial^{j})\overset{\partial|y}{\triangleright}g(y^{k})  &
=\sum_{a}\mathcal{W}(e_{a})\otimes\big\langle f^{a},g(y^{k})_{(1)}%
\big\rangle\,g(y^{k})_{(2)}\nonumber\\
&  =\sum_{a}\mathcal{W}(e_{a})\otimes\varepsilon\big (f^{a}\triangleleft
g(y^{k})_{(1)}\big )g(y^{k})_{(2)}\nonumber\\
&  =\sum_{a}\big (g(x^{i})_{(1)}\overset{x}{\circledast}\mathcal{W}%
(e_{a})\big )\otimes\varepsilon\left(  f^{a}\right)  g(y^{k})_{(2)}\nonumber\\
&  =g(x^{i})_{(1)}\otimes g(y^{k})_{(2)}=g(x^{i}\oplus y^{k}),
\end{align}
and%
\begin{align}
g(y^{k})\overset{y|\partial}{\triangleleft}\exp(\partial^{j}|x^{i})  &
=\sum_{a}g(y^{k})_{(1)}\big\langle g(y^{k})_{(2)},f^{a}\big\rangle^{\prime
}\otimes\mathcal{W}(e_{a})\nonumber\\
&  =\sum_{a}g(y^{k})_{(1)}\,\varepsilon\big (g(y^{k})_{(2)}\triangleright
f^{a}\big )^{\prime}\otimes\mathcal{W}(e_{a})\nonumber\\
&  =\sum_{a}g(y^{k})_{(1)}\,\varepsilon\left(  f^{a}\right)  \otimes
\big (\mathcal{W}(e_{a})\circledast g(x^{i})_{(2)}\big )\nonumber\\
&  =g(y^{k})_{(1)}\otimes g(x^{i})_{(2)}=g(y^{k}\oplus x^{i}).
\end{align}
For the first step we applied (\ref{CorActN}) and the second equality is the
definition of the pairing. The third equality uses (\ref{EigGl1N}) and
(\ref{EigGl2}). The fourth equality results from (\ref{NorExpN}).

If we want to apply the formulae in (\ref{q-TayN}) we need to know the
explicit form of the q-exponential. As it was pointed out above the
q-exponential is completely determined as soon as we know two bases being dual
to each other. For this reason we consider a basis of normally ordered
monomials and try to find the corresponding dual basis. As will become clear
later on, this task requires to know the values of the dual pairing on
normally ordered monomials.

A short glance at (\ref{DefPar1N}) and (\ref{DefPar2N}) tells us that the dual
pairings depend on the action of partial derivatives on quantum spaces. Due to
this fact we first start with a discussion of covariant differential calculus
on q-deformed Minkowski space, for which we have
\begin{align}
\partial^{\mu}X^{\nu}  &  =\eta^{\mu\nu}+q^{-2}(\hat{R}_{II}^{-1})_{\rho
\sigma}^{\mu\nu}X^{\rho}\partial^{\sigma},\nonumber\\
\hat{\partial}^{\mu}X^{\nu}  &  =\eta^{\mu\nu}+q^{2}(\hat{R}_{II})_{\rho
\sigma}^{\mu\nu}X^{\rho}\hat{\partial}^{\sigma},\quad\mu,\nu\in\{\pm,0,3\}.
\label{ParQ-Min}%
\end{align}
Notice that $\hat{R}_{II}$ stands for one of the two R-matrices of the
q-deformed Lorentz-algebra \cite{LSW94} and $\eta^{\mu\nu}$ denotes the
corresponding quantum metric. In the following we focus attention on the
second relation in (\ref{ParQ-Min}). It implies the identities
\begin{align}
\hat{\partial}^{\mu}\hat{r}^{2}  &  =q^{-1}\lambda_{+}X^{\mu}+q^{2}\hat{r}%
^{2}\hat{\partial}^{\mu},\nonumber\\
\hat{\partial}^{2}X^{\mu}  &  =q^{-1}\lambda_{+}\hat{\partial}^{\mu}%
+q^{2}X^{\mu}\hat{\partial}^{2},\nonumber\\
\hat{\partial}^{2}\hat{r}^{2}  &  =q\lambda_{+}^{3}-q^{3}\lambda_{+}%
(X\circ\hat{\partial})+q^{4}\hat{r}^{2}\hat{\partial}^{2},\nonumber\\
(X\circ\hat{\partial})\hat{r}^{2}  &  =q^{2}\hat{r}^{2}(X\circ\hat{\partial
})-\lambda_{+}\hat{r}^{2}, \label{VerRegMin}%
\end{align}
where%
\begin{align}
\hat{r}^{2}  &  \equiv\eta_{\mu\nu}X^{\mu}X^{\nu}\nonumber\\
&  =-X^{0}X^{0}+X^{3}X^{3}-qX^{+}X^{-}-q^{-1}X^{-}X^{+}\nonumber\\
&  =\lambda_{+}X^{+}X^{-}-q^{-1}\lambda_{+}X^{0}X^{3/0}-q^{-2}X^{3/0}X^{3/0},
\label{Defr2N}%
\end{align}
and%
\begin{equation}
\hat{\partial}^{2}\equiv\eta_{\mu\nu}\hat{\partial}^{\mu}\hat{\partial}^{\nu
},\quad X\circ\hat{\partial}\equiv\eta_{\mu\nu}X^{\mu}\hat{\partial}^{\nu}.
\end{equation}

Next, we would like to derive expressions for the action of partial
derivatives on normally ordered monomials of the form%
\begin{equation}
(\hat{r}^{2})^{n_{r}}(X^{+})^{n_{+}}(X^{3/0})^{n_{3/0}-n_{r}}(X^{-})^{n_{-}%
},\quad n_{\mu}\in\mathbb{N}_{0}. \label{MinBasRa}%
\end{equation}
Let us notice that $X^{+}$, $X^{-}$, and $X^{3/0}$ are q-analogs of light-cone
coordinates, whereas $\hat{r}^{2}$ denotes the square of the Minkowski length.
The coordinate $X^{3/0}$ is related to the space coordinate $X^{3}$ and the
time element $X^{0}$ by $X^{3/0}\equiv X^{3}-X^{0}.$ In appendix
\ref{AppTrans} it is shown that the monomials in (\ref{MinBasRa})\ establish a
basis, which can be transformed into a more usual one with $\hat{r}^{2}$ being
substituted by the time coordinate $X^{0}$. The reason for using the basis in
(\ref{MinBasRa}) lies in the fact that it helps to simplify\ our calculations.

To get left actions of partial derivatives we repeatedly apply the commutation
relations (\ref{ParQ-Min}) and (\ref{VerRegMin}) to a product of a partial
derivative with a normally ordered monomial of the form (\ref{MinBasRa}),
until we obtain an expression with all partial derivatives standing to the
right of all quantum space coordinates, i.e.%
\begin{gather}
\hat{\partial}^{\mu}(\hat{r}^{2})^{n_{r}}(X^{+})^{n_{+}}(X^{3/0}%
)^{n_{3/0}-\,n_{r}}(X^{-})^{n_{-}}\nonumber\\
=\big (\hat{\partial}_{(1)}^{\mu}\,\bar{\triangleright}\,(\hat{r}^{2})^{n_{r}%
}(X^{+})^{n_{+}}(X^{3/0})^{n_{3/0}-\,n_{r}}(X^{-})^{n_{-}}\big )\hat{\partial
}_{(2)}^{\mu}.\label{VerRelAblMon}%
\end{gather}
Then we take the counit of partial derivatives appearing on the right-hand
side of the last expression in (\ref{VerRelAblMon}). This yields an expression
for the left action of $\hat{\partial}^{\mu}$, since it holds%
\begin{gather}
\big (\partial_{(1)}^{\mu}\triangleright(\hat{r}^{2})^{n_{r}}(X^{+})^{n_{+}%
}(X^{3/0})^{n_{3/0}-\,n_{r}}(X^{-})^{n_{-}}\big )\varepsilon(\partial
_{(2)}^{\mu})\nonumber\\
=\partial^{\mu}\triangleright(\hat{r}^{2})^{n_{r}}(X^{+})^{n_{+}}%
(X^{3/0})^{n_{3/0}-\,n_{r}}(X^{-})^{n_{-}}.
\end{gather}
Recalling the definitions in\ (\ref{DefParAcN}) and the considerations about
q-translations [see the discussion of (\ref{OpRep1})-(\ref{OpRep2})] our
results should enable us to read off operator expressions for the action of
partial derivatives on commutative functions. Proceeding in this manner and
fixing the algebra isomorphism in Eq. (\ref{AlgIsoN}) by
\begin{equation}
\mathcal{W}((r^{2})^{n_{r}}(x^{+})^{n_{+}}(x^{3/0})^{n_{3/0}}(x^{-})^{n_{-}%
})=(\hat{r}^{2})^{n_{r}}(X^{+})^{n_{+}}(X^{3/0})^{n_{3/0}}(X^{-})^{n_{-}%
},\label{AlgIsoMin}%
\end{equation}
we finally obtain
\begin{align}
&  \hspace{-0.16in}\hat{\partial}^{3/0}\,\bar{\triangleright}\,f(r^{2}%
,x^{+},x^{3/0},x^{-})=q^{-1}\lambda_{+}x^{3/0}D_{q^{2}}^{r^{2}}f(q^{2}%
x^{+}),\\[0.16in]
&  \hspace{-0.16in}\hat{\partial}^{+}\,\bar{\triangleright}\,f(r^{2}%
,x^{+},x^{3/0},x^{-})=-qD_{q^{2}}^{-}f(q^{2}r^{2})+q^{-1}\lambda_{+}%
x^{+}D_{q^{2}}^{r^{2}}f,\\[0.16in]
&  \hspace{-0.16in}\hat{\partial}^{-}\,\bar{\triangleright}\,f(r^{2}%
,x^{+},x^{3/0},x^{-})=-q^{-1}D_{q^{2}}^{+}f\nonumber\\
\, &  \quad+q^{-1}\lambda_{+}x^{-}D_{q^{2}}^{r^{2}}f(q^{2}x^{+},q^{2}%
x^{3/0})\nonumber\\
&  \quad+\,q^{-2}\lambda(x^{3/0})^{2}D_{q^{2}}^{+}D_{q^{2}}^{r^{2}}%
f(q^{2}x^{+},q^{2}x^{-})\nonumber\\
&  \quad-\,q^{-1}\lambda^{2}(x^{3/0})^{2}x^{-}D_{q^{2}}^{+}D_{q^{2}}%
^{-}D_{q^{2}}^{r^{2}}f(q^{2}x^{+}),\\[0.16in]
&  \hspace{-0.16in}\hat{\partial}^{2}\,\bar{\triangleright}\,f(r^{2}%
,x^{+},x^{3/0},x^{-})=q^{4}\lambda_{+}^{2}(D_{q^{2}}^{r^{2}})^{2}f(q^{2}%
x^{+},q^{2}x^{3/0},q^{2}x^{-})\nonumber\\
\, &  \quad+\lambda_{+}^{2}r^{-2}(D_{q^{2}}^{r^{2}})^{2}r^{2}f(q^{2}%
x^{+},q^{2}x^{3/0},q^{2}x^{-})\nonumber\\
&  \quad+\,(q^{-1}\lambda_{+})^{2}x^{-}D_{q^{2}}^{-}D_{q^{2}}^{r^{2}}%
f(q^{2}x^{+},q^{2}x^{3/0})\nonumber\\
&  \quad+\,(q^{-1}\lambda_{+})^{2}x^{3/0}D_{q^{2}}^{3/0}D_{q^{2}}^{r^{2}%
}f(q^{2}x^{+})\nonumber\\
&  \quad-\,\lambda_{+}D_{q^{2}}^{+}D_{q^{2}}^{-}f(q^{2}r^{2})+(q^{-1}%
\lambda_{+})^{2}x^{+}D_{q^{2}}^{+}D_{q^{2}}^{r^{2}}f.
\end{align}
One can even show that the following more general formulae hold:
\begin{align}
&  \hspace{-0.16in}(\hat{\partial}^{3/0})^{m_{3/0}}\,\bar{\triangleright
}\,f(r^{2},x^{+},x^{3/0},x^{-})=(q^{-1}\lambda_{+}x^{3/0})^{m_{3/0}}(D_{q^{2}%
}^{r^{2}})^{m_{3/0}}f(q^{2m_{3/0}}x^{+}),\label{PotPD3N}\\[0.16in]
&  \hspace{-0.16in}(\hat{\partial}^{+})^{m_{+}}\,\bar{\triangleright}%
\,f(r^{2},x^{+},x^{3/0},x^{-})=\sum_{k=0}^{m_{+}}(-q)^{m_{+}}(-q^{-2}%
\lambda_{+}x^{+})^{k}%
\genfrac{[}{]}{0pt}{}{m_{+}}{k}%
_{q^{2}}\nonumber\\
&  \qquad\times(D_{q^{2}}^{-})^{m_{+}-\,k}(D_{q^{2}}^{r^{2}})^{k}%
f(q^{2(m_{+}-\,k)}r^{2}),\\[0.16in]
&  \hspace{-0.16in}(\hat{\partial}^{-})^{m_{-}}\,\bar{\triangleright}%
\,f(r^{2},x^{+},x^{3/0},x^{-})=\sum_{i=0}^{m_{-}}\sum_{j=0}^{i}\sum
_{k=0}^{i-j}(-1)^{m_{-}-\,k}q^{-2m_{-}-\,j}\lambda_{+}^{i+j-k}\lambda
^{j+2k}\nonumber\\
&  \qquad\times q^{-k(k+1)-2i^{2}-\,4j^{2}-\,2k^{2}+\,4ij-\,6kj+2ki-m_{-}%
(m_{-}-\,2i+2j-2k)}\nonumber\\
&  \qquad\times%
\genfrac{[}{]}{0pt}{}{i}{j}%
_{q^{-2}}%
\genfrac{[}{]}{0pt}{}{i-j}{k}%
_{q^{-2}}%
\genfrac{[}{]}{0pt}{}{m_{-}}{i}%
_{q^{-2}}\nonumber\\
&  \qquad\times(x^{3/0})^{2(j+k)}(x^{-})^{i-j}\,(D_{q^{-2}}^{+})^{m_{-}%
-\,i+j+k}(D_{q^{-2}}^{-})^{k}(D_{q^{-2}}^{r^{2}})^{i}\nonumber\\
&  \qquad\times f(q^{2i}r^{2},q^{2(m_{-}+\,j+k)}x^{+},q^{2(i-j-k)}%
x^{3/0},q^{2(j+k)}x^{-}),\\[0.16in]
&  \hspace{-0.16in}(\hat{\partial}^{2})^{m_{r}}\,\bar{\triangleright}%
\,f(r^{2},x^{+},x^{3/0},x^{-})=\sum_{i=0}^{m_{r}}\sum_{j=0}^{m_{r}-\,i}%
\sum_{k=0}^{m_{r}-\,i-j}\sum_{l=0}^{i}(q^{-1}\lambda_{+})^{i+l+2(j+k)}%
\nonumber\\
&  \qquad\times(-q^{-2(m_{r}-\,i)+1})^{i-l}\frac{[[i+j+k]]_{q^{-2}}%
!}{[[i]]_{q^{-2}}![[j]]_{q^{-2}}![[k]]_{q^{-2}}!}\,%
\genfrac{[}{]}{0pt}{}{i}{l}%
_{q^{2}}%
\genfrac{[}{]}{0pt}{}{m_{r}}{i+j+k}%
_{q^{-2}}\nonumber\\
&  \qquad\times(x^{+})^{l}(x^{3/0})^{j}(D_{q^{2}}^{3/0})^{j}(D_{q^{2}}%
^{+})^{i}(D_{q^{2}}^{-})^{i-l}(x^{-})^{k}(D_{q^{2}}^{-})^{k}(D_{q^{2}}^{r^{2}%
})^{j+k+l}\nonumber\\
&  \qquad\times\Big [(q\lambda_{+})^{2}\big (q^{2}(D_{q^{2}}^{r^{2}}%
)^{2}+q^{-2}r^{-2}(D_{q^{2}}^{r^{2}})^{2}r^{2}\big )\Big ]^{m_{r}%
-\,i-j-k}\nonumber\\
&  \qquad\times f(q^{2(i-l)}r^{2},q^{2(m_{r}-\,i)}x^{+},q^{2(m_{r}%
-\,i-j)}x^{3/0},q^{2(m_{r}-\,i-j-k)}x^{-}).\label{PD3dimEndN}%
\end{align}
Let us mention that formula (\ref{PotPD3N}) also holds for negative values of
$m_{3/0}$ if we set
\begin{equation}
(D_{q^{a}})^{-1}x^{n}\equiv\frac{1}{[[n+1]]_{q^{a}}}x^{n+1},
\end{equation}
or more general
\begin{equation}
(D_{q^{a}})^{-1}f(x)\equiv-(1-q^{a})\sum_{k=1}^{\infty}(q^{-ak}x)f(q^{-ak}%
x),\quad q>1.
\end{equation}

Now, we come to the calculation of the pairing in (\ref{DefPar1N}). For our
purposes it is sufficient to know its values on normally ordered monomials of
the form
\begin{align}
\hat{\partial}^{\,\underline{n}}  &  \equiv(\hat{\partial}^{-})^{n_{-}}%
(\hat{\partial}^{3/0})^{n_{3/0}-\,n_{r}}(\hat{\partial}^{+})^{n_{+}}%
(\hat{\partial}^{2})^{n_{r}}\in\mathcal{A}_{q}^{\ast},\nonumber\\
X^{\underline{m}}  &  \equiv(X^{+})^{m_{+}}(X^{3/0})^{m_{3/0}-\,m_{r}}%
(X^{-})^{m_{-}}(\hat{r}^{2})^{m_{r}}\in\mathcal{A}_{q},
\end{align}
i.e. we limit ourselves to the evaluation of%
\begin{equation}
\big \langle\hat{\partial}^{\,\underline{n}},X^{\underline{m}}%
\big \rangle_{\bar{L},R}=\varepsilon(\hat{\partial}^{\,\underline{n}}%
\,\bar{\triangleright}\,X^{\underline{m}}). \label{ParMon}%
\end{equation}
This way, we first apply the actions in (\ref{PotPD3N})-(\ref{PD3dimEndN}) in
succession and then take the counit. After some tedious steps we obtain the
expression%
\begin{align}
\big \langle\hat{\partial}^{\,\underline{n}},X^{\underline{m}}%
\big \rangle_{\bar{L},R}  &  =\sum_{i=0}^{n_{r}}\sum_{j=0}^{n_{r}-\,i}%
\sum_{k=0}^{n_{r}-\,i-j}\sum_{l=0}^{i}(q^{-1}\lambda_{+})^{2(i+j+k)+m_{r}%
-\,n_{r}}(-q)^{m_{-}-\,n_{-}}\nonumber\\
&  \times q^{i(i-1)+j(j-1)+k(k-1)+2l(i-l)+2m_{-}m_{r}+2n_{-}(n_{3/0}-\,n_{r}%
)}\nonumber\\
&  \times q^{2m_{+}(n_{r}-\,i)+2(m_{3/0}-\,m_{r})(n_{r}-\,i-j)-2m_{-}%
(j+k+l)}\nonumber\\
&  \times\lbrack\lbrack i+j+k]]_{q^{-2}}!\,[[m_{r}-(n_{r}-i-j-k)]]_{q^{2}%
}!\nonumber\\
&  \times\lbrack\lbrack m_{-}]]_{q^{2}}!\,[[n_{-}]]_{q^{2}}!\,%
\genfrac{[}{]}{0pt}{}{i}{l}%
_{q^{2}}%
\genfrac{[}{]}{0pt}{}{n_{r}}{i+j+k}%
_{q^{-2}}%
\genfrac{[}{]}{0pt}{}{n_{+}}{m_{-}-(i-l)}%
_{q^{2}}\nonumber\\
&  \times%
\genfrac{[}{]}{0pt}{}{m_{+}}{i}%
_{q^{2}}%
\genfrac{[}{]}{0pt}{}{m_{3/0}-m_{r}}{j}%
_{q^{2}}%
\genfrac{[}{]}{0pt}{}{m_{-}}{k}%
_{q^{2}}\nonumber\\
&  \times\delta_{m_{3/0}-\,m_{r},-n_{3/0}+\,n_{r}}\,\delta_{m_{r}%
+\,m_{-},n_{+}+\,n_{3/0}}\,\delta_{m_{+}-\,m_{-},n_{-}-\,n_{+}}.
\label{PaaMinExpN}%
\end{align}
It should be noticed that (\ref{ParMon}) does not provide the only possibility
for a pairing between partial derivatives and quantum space coordinates. For
this reason we introduced the labels $\bar{L}$ and $R$ to distinguish the
pairing in (\ref{ParMon}) from other versions. For the details we refer the
reader to Ref. \cite{Wac03}.

Now, we are in a position to find the basis which is dual to a basis of
normally ordered monomials. First of all, it is not very difficult to check
that the expression in (\ref{PaaMinExpN}) does not vanish iff
\begin{align}
m_{\pm}  &  =n_{\mp}-v,\nonumber\\
m_{r}  &  =n_{3/0}+v,\nonumber\\
m_{3/0}-m_{r}  &  =-(n_{3/0}-n_{r}),\nonumber\\
-n_{3/0}  &  \leq v\leq\min(n_{+},n_{-}). \label{Schrank}%
\end{align}
In other words, this means%
\begin{gather}
\big \langle\hat{\partial}^{\,\underline{n}},X^{\underline{m}}%
\big \rangle_{\bar{L},R}\neq0\nonumber\\
\Leftrightarrow\,X^{\underline{m}}=(X^{+})^{n_{-}-\,v}(X^{3/0})^{-(n_{3/0}%
-\,n_{r})}(X^{-})^{n_{+}-\,v}(\hat{r}^{2})^{n_{3/0}+\,v}\nonumber\\
\text{with}\quad-n_{3/0}\leq v\leq\min(n_{+},n_{-}).
\end{gather}
The last assertion implies that the element being dual to $\hat{\partial
}^{\,\underline{n}}$ has to be of the form
\begin{align}
f^{\underline{n}}(X^{\mu})  &  \equiv\sum_{-n_{3/0}\leq v\leq\min(n_{+}%
,n_{-})}f_{v}^{\underline{n}}\nonumber\\
&  \times(X^{+})^{n_{-}-\,v}(X^{3/0})^{-(n_{3/0}-\,n_{r})}(X^{-})^{n_{+}%
-\,v}(r^{2})^{n_{3/0}+\,v}, \label{CoefExp}%
\end{align}
where it remains to determine the unknown coefficients $f_{v}^{\underline{n}}$.

Next, we try to find a system of equations for the unknown coefficients
$f_{v}^{\underline{n}}.$ Since $f^{\underline{n}}(X^{\mu})$ and $\hat
{\partial}^{\,\underline{k}}$ denote elements of two dually paired bases they
have to be subject to
\begin{equation}
\big \langle\hat{\partial}^{\,\underline{k}},f^{\underline{n}}(X^{\mu
})\big \rangle_{\bar{L},R}=\delta_{\underline{k},\underline{n}}, \label{BesGl}%
\end{equation}
where%
\begin{equation}
\delta_{\underline{k},\underline{n}}=\delta_{k_{+},n_{+}}\delta_{k_{-},n_{-}%
}\delta_{k_{3/0},n_{3/0}}\delta_{k_{r},n_{r}}.
\end{equation}
Substituting the expression in (\ref{CoefExp}) for $f^{\underline{n}}(X^{\mu
})$, the relation in (\ref{BesGl})\ gives
\begin{equation}
\sum_{-n_{3/0}\leq v\leq\min(n_{+},n_{-})}f_{v}^{\underline{n}}\cdot
\big \langle\hat{\partial}^{\,\underline{k}},X^{\underline{n}-(v,-v,-v,v)}%
\big \rangle_{\bar{L},R}=\delta_{\underline{k},\underline{n}},
\end{equation}
where it is understood that%
\begin{align}
\underline{n}-(v,-v,-v,v)  &  =(n_{+}-v,n_{3/0}+v,n_{r}+v,n_{-}-v),\\
(  &  \equiv(n_{+}^{\prime},n_{3/0}^{\prime},n_{r}^{\prime},n_{-}^{\prime
})=\underline{n}^{\prime}\,).\nonumber
\end{align}
If \underline{$k$} is specified according to
\begin{equation}
\underline{k}=\underline{n}-(v^{\prime},-v^{\prime},-v^{\prime},v^{\prime
}),\quad v^{\prime}=-n_{3/0},\ldots,\min(n_{+},n_{-}),
\end{equation}
we get a system of $\min(n_{+},n_{-})+n_{3/0}+1$ equations for the same number
of coefficients $f_{v}^{\underline{n}}$, i.e.
\begin{equation}
\sum_{%
\genfrac{}{}{0pt}{}{-n_{3/0}+\max(0,v^{\prime})\leq v,}{v\leq\min(n_{+}%
,n_{-})+\min(0,v^{\prime})}%
}f_{v}^{\underline{n}}\cdot\big \langle v^{\prime},v\big \rangle_{\bar{L}%
,R}^{\underline{n}}=\delta_{0}^{v^{\prime}},\quad v^{\prime}=-n_{3/0}%
,\ldots,\min(n_{+},n_{-}), \label{GlSys}%
\end{equation}
where, for brevity, we have introduced
\begin{equation}
\big \langle v^{\prime},v\big \rangle_{\bar{L},R}^{\underline{n}}%
\equiv\big \langle\hat{\partial}^{\,\underline{n}-(v^{\prime},-v^{\prime
},-v^{\prime},v^{\prime})},X^{\underline{n}-(v,-v,-v,v)}\big \rangle_{\bar
{L},R}. \label{NeuPaar}%
\end{equation}
In (\ref{GlSys}) we were able to restrict the range for the summation variable
$v$. The reason for this becomes quite clear, when we take a short look at
(\ref{Schrank}) telling us that the pairing in (\ref{NeuPaar}) vanishes for\
\begin{equation}
-n_{3/0}\leq v-v^{\prime}\leq\min(n_{+},n_{-}).
\end{equation}

The above system of equations can be solved by applying standard techniques
such as the Gaussian elimination method. This task can be done best with the
help of computer algebra systems. In doing so, we will obtain expressions for
the $f_{v}^{\underline{n}}$ in terms of dual pairings. Inserting the solutions
to the $f_{v}^{\underline{n}}$ into (\ref{CoefExp}) will provide us with
explicit formulae for the basis elements $f^{\underline{n}}(X^{\mu})$.
Obviously, the exponential corresponding to the pairing in (\ref{ParMon}) then
becomes%
\begin{equation}
\exp(x^{\mu}|\hat{\partial}^{\nu})_{R,\bar{L}}=\sum_{\underline{n}=0}^{\infty
}f^{\underline{n}}(x^{\mu})\otimes(\hat{\partial}^{-})^{n_{-}}(\hat{\partial
}^{3/0})^{n_{3/0}-\,n_{r}}(\hat{\partial}^{+})^{n_{+}}(\hat{\partial}%
^{2})^{n_{r}}. \label{AnsExp}%
\end{equation}
Knowing the explicit form of the q-deformed exponential and that for the
action of partial derivatives, we should now have everything together for
calculating q-deformed translations by applying the formula%
\begin{equation}
g(x^{\mu}\oplus_{L}y^{\nu})=\exp(x^{\mu}|\hat{\partial}^{\rho})_{R,\bar{L}%
}\,\overset{\partial|y}{\bar{\triangleright}}\,g(y^{\nu}).
\end{equation}

Next, we turn to the antipode on q-deformed Minkowski space. On coordinates it
takes the form\ \cite{BW01}%
\begin{align}
S_{L}(X^{3/0})=  &  -\Lambda^{1/2}\sigma^{2}X^{3/0}-q^{-3/2}\lambda_{+}%
^{1/2}\lambda\Lambda^{1/2}S^{1}X^{+},\nonumber\\
S_{L}(X^{+})=  &  -\Lambda^{1/2}\tau^{1}(\tau^{3})^{1/2}X^{+}-q^{3/2}%
\lambda_{+}^{-1/2}\lambda\Lambda^{1/2}T^{2}(\tau^{3})^{1/2}X^{3/0},\nonumber\\
S_{L}(X^{-})=  &  -\Lambda^{1/2}\sigma^{2}(\tau^{3})^{-1/2}X^{-}%
-q^{-1/2}\lambda_{+}^{1/2}\lambda\Lambda^{1/2}(\tau^{3})^{-1/2}S^{1}%
X^{0}\nonumber\\
&  +\,q^{-2}\lambda^{2}\Lambda^{1/2}(\tau^{3})^{-1/2}S^{1}T^{-}X^{+}%
\nonumber\\
&  +\,q^{-5/2}\lambda_{+}^{-1/2}\lambda\Lambda^{1/2}(\tau^{3})^{-1/2}%
(\sigma^{2}T^{-}-q^{3}S^{1})X^{3/0},\nonumber\\
S_{L}(X^{0})=  &  -\Lambda^{1/2}\tau^{1}X^{0}-q^{5/2}\lambda_{+}^{-1/2}%
\lambda\Lambda^{1/2}T^{2}X^{-}\nonumber\\
&  +\,q^{-3/2}\lambda_{+}^{-1/2}\lambda\Lambda^{1/2}(\tau^{1}T^{-}%
+qS^{1})X^{+}\nonumber\\
&  +\,\lambda_{+}^{-1}\Lambda^{1/2}(q(\sigma^{2}-\tau^{1})+\lambda^{2}%
T^{2}T^{-})X^{3/0}.
\end{align}
Notice that $T^{-},$ $T^{2},$ $S^{1},$ $\sigma^{2},$ $\tau^{1},$ and $\tau
^{3}$ are generators of the q-deformed Lorentz algebra, while $\Lambda$
denotes a scaling operator. In complete analogy to the considerations of the
previous sections we derived from the above relations the following formula:
\begin{align}
&  \hspace{-0.16in}S_{L}((X^{+})^{n_{+}}(X^{3/0})^{n_{3/0}}(X^{-})^{n_{-}%
})\nonumber\\
=\,  &  (-1)^{n_{+}+n_{3/0}+n_{-}}q^{n_{+}(n_{+}-1)+n_{3/0}(n_{3/0}%
-1)+n_{-}(n_{-}-1)}\nonumber\\
&  \times\sum_{k=0}^{n_{+}}(-\lambda_{+}^{-1}\lambda)^{k}q^{k(4k+1)-2k(n_{+}%
-\,n_{3/0}+\,n_{-})}[[k]]_{q^{2}}!%
\genfrac{[}{]}{0pt}{}{n_{+}}{k}%
_{q^{2}}%
\genfrac{[}{]}{0pt}{}{n_{-}}{k}%
_{q^{2}}\nonumber\\
&  \qquad\times m_{\mathcal{H}\otimes A_{q}}((X_{r}^{-})^{n_{-}-\,k}%
(X_{r}^{3/0})^{n_{3/0}+2k}(X_{r}^{+})^{n_{+}-\,k}).
\end{align}

Now, we want to extend this result to monomials depending on\ $\hat{r}^{2}.$
Towards this end we need the quantities
\begin{equation}
\hat{r}_{r}^{2}\equiv\eta_{\mu\nu}X_{r}^{\mu}X_{r}^{\nu}=\Lambda^{-1}%
\otimes\hat{r}^{2},
\end{equation}
and
\begin{equation}
\hat{r}_{l}^{2}\equiv\eta_{\mu\nu}X_{l}^{\mu}X_{l}^{\nu}=\hat{r}^{2}\otimes1,
\end{equation}
which fulfill
\begin{equation}
r_{r}^{2}X_{l}^{\mu}=q^{2}X_{l}^{\mu}r_{r}^{2},\quad r_{r}^{2}r_{l}^{2}%
=q^{4}r_{l}^{2}r_{r}^{2}.
\end{equation}
A direct calculation using the definition of $\hat{r}^{2}$ together with the
antipodes on coordinates shows us that%
\begin{equation}
S_{L}(\hat{r}^{2})=q^{-2}\Lambda^{-1}\hat{r}^{2}.
\end{equation}
With these identities at hand one can check that
\begin{align}
&  \hspace{-0.16in}S_{L}((\hat{r}^{2})^{n_{r}}(X^{+})^{n_{+}}(X^{3/0}%
)^{n_{3/0}}(X^{-})^{n_{-}})\nonumber\\
=\,  &  (-1)^{n_{+}+\,n_{3/0}+n_{-}}q^{n_{+}(n_{+}-1)+n_{3/0}(n_{3/0}%
-1)+n_{-}(n_{-}-1)}\nonumber\\
&  \times q^{2n_{r}(n_{+}+\,n_{3/0}+n_{-}+\,n_{r}-2)}\nonumber\\
&  \times\sum_{k=0}^{n_{+}}(-\lambda\lambda_{+}^{-1})^{k}q^{k(4k+1)-2k(n_{+}%
-\,n_{3/0}+n_{-})}[[k]]_{q^{2}}!%
\genfrac{[}{]}{0pt}{}{n_{+}}{k}%
_{q^{2}}%
\genfrac{[}{]}{0pt}{}{n_{-}}{k}%
_{q^{2}}\nonumber\\
&  \qquad\times m_{\mathcal{H}\otimes A_{q}}((X_{r}^{-})^{n_{-}-\,k}%
(X_{r}^{3/0})^{n_{3/0}+2k}(X_{r}^{+})^{n_{+}-\,k}(\hat{r}_{r}^{2})^{n_{r}}),
\end{align}
which, in turn, leads to%
\begin{align}
&  \hspace{-0.16in}\hat{U}(f(\ominus_{L}\,x^{\mu}))=\sum_{k=0}^{\infty
}(-q\lambda\lambda_{+}^{-1})^{k}\,\frac{q^{4k^{2}}(x^{3/0})^{2k}}%
{[[k]]_{q^{2}}!}\nonumber\\
&  \qquad\times(D_{q^{2}}^{+}D_{q^{2}}^{-})^{k}\,q^{\hat{n}_{+}^{2}+\,\hat
{n}_{3/0}^{2}+\,\hat{n}_{-}^{2}+\,2\hat{n}_{r}(\hat{n}_{+}+\,\hat{n}%
_{3/0}+\hat{n}_{-}+\,\hat{n}_{r})}\nonumber\\
&  \qquad\times f(q^{-4}r^{2},-q^{-2k-1}x^{+},-q^{2k-1}x^{3/0},-q^{-2k-1}%
x^{-}).
\end{align}
Again the operator $\hat{U}$ transforms functions referring to ordering
$\hat{r}^{2}X^{+}X^{3/0}X^{-}$ into those referring\ to reversed ordering. Its
explicit form was already given in Ref. \cite{BW01}.

In complete analogy to the previous sections we can assign different types of
Hopf structures to q-deformed Minkowski space. Again, from\ these Hopf
structures arise different versions of q-translations, which are related to
each other by the crossing-symmetries%
\begin{align}
f(x^{\mu}\oplus_{L}y^{\nu})  &  \overset{{%
\genfrac{}{}{0pt}{}{\pm}{q}%
}{%
\genfrac{}{}{0pt}{}{\rightarrow}{\rightarrow}%
}{%
\genfrac{}{}{0pt}{}{\mp}{1/q}%
}}{\longleftrightarrow}f(x^{\mu}\,\widetilde{\oplus}_{\bar{L}}\,y^{\nu
}),\\[0.16in]
f(x^{\mu}\oplus_{L}y^{\nu})  &  \overset{+\leftrightarrow-}%
{\longleftrightarrow}f(x^{\mu}\oplus_{\bar{R}}y^{\nu}),\nonumber\\
f(x^{\mu}\,\widetilde{\oplus}_{R}\,y^{\nu})  &  \overset{+\leftrightarrow
-}{\longleftrightarrow}f(x^{\mu}\,\widetilde{\oplus}_{\bar{L}}\,y^{\nu}),
\end{align}
and
\begin{align}
\hat{U}(f(\ominus_{L}\,x^{\mu}))  &  \overset{{%
\genfrac{}{}{0pt}{}{\pm}{q}%
}{%
\genfrac{}{}{0pt}{}{\rightarrow}{\rightarrow}%
}{%
\genfrac{}{}{0pt}{}{\mp}{1/q}%
}}{\longleftrightarrow}\hat{U}^{-1}(f(\widetilde{\ominus}_{\bar{L}}\,x^{\mu
})),\\[0.16in]
\hat{U}(f(\ominus_{L}\,x^{\mu}))  &  \overset{+\leftrightarrow-}%
{\longleftrightarrow}\hat{U}(f(\ominus_{\bar{R}}\,x^{\mu})),\nonumber\\
\hat{U}^{-1}(f(\widetilde{\ominus}_{R}\,x^{\mu}))  &  \overset
{+\leftrightarrow-}{\longleftrightarrow}\hat{U}^{-1}(f(\widetilde{\ominus
}_{\bar{L}}\,x^{\mu})),
\end{align}
where $\overset{{%
\genfrac{}{}{0pt}{}{\pm}{q}%
}{%
\genfrac{}{}{0pt}{}{\rightarrow}{\rightarrow}%
}{%
\genfrac{}{}{0pt}{}{\mp}{1/q}%
}}{\longleftrightarrow}$ and $\overset{+\leftrightarrow-}{\longleftrightarrow
}$ have the same meaning as in Sec. \ref{Kap2}.

\section{Conclusion}

Let us make a few comments on our results. The aim of our program is to
provide us with a q-deformed version of classical analysis. Translations on
quantum spaces are an important ingredient of q-analysis. This article is
devoted to the explicit calculation of q-deformed translations on quantum
spaces of physical interest, i.e. Manin plane, q-deformed Euclidean space in
three or four dimensions, and q-deformed Minkowski space. In some sense, our
results can be viewed as q-analogs of classical Taylor rules. These
q-deformed-Taylor rules show a structure being similar to that of their
classical counterparts. However, in contrast to the classical case they are
not formulated by means of ordinary factorials and derivatives. Instead we
have to deal with q-factorials and Jackson derivatives. Furthermore, one can
see that q-Taylor rules are often modified by non-classical terms, which
depend on the parameter $\lambda=q-q^{-1}.$ These observations should tell us
that our results tend to their classical counterparts as $q\rightarrow1.$

Next, we would like to discuss that q-deformed translations show a number of
properties which are very similar to those fulfilled by\ classical
translations. As we know, classical translations form a group and every group
is a Hopf algebra. Important for us is the fact that\ q-deformed translations
are based on a Hopf structure. It is not very difficult to convince oneself
that the Hopf algebra axioms%
\begin{gather}
(\Delta\otimes\text{id})\circ\Delta=(\text{id}\otimes\Delta)\circ
\Delta,\nonumber\\
(\text{id}\otimes\varepsilon)\circ\Delta=(\varepsilon\otimes\text{id}%
)\circ\Delta=\text{id},\nonumber\\
m\circ(S\otimes\text{id})\circ\Delta=\varepsilon=m\circ(\text{id}\otimes
S)\circ\Delta, \label{HopfAxiom0}%
\end{gather}
imply for q-deformed translations that%
\begin{align}
f((x^{i}\oplus_{A}y^{j})\oplus_{A}(\ominus_{A}\,z^{k}))  &  =f(x^{i}\oplus
_{A}(y^{j}\oplus_{A}(\ominus_{A}\,z^{k}))),\nonumber\\
f(0\oplus_{A}x^{i})  &  =f(x^{i}\oplus_{A}0)=f(x^{i}),\nonumber\\
f((\ominus_{A}\,x^{i})\oplus_{A}x^{j})  &  =f(x^{i}\oplus_{A}(\ominus_{A}%
x^{j}))=f(0). \label{HopfAxiom}%
\end{align}
For a correct understanding of the relations in (\ref{HopfAxiom}) one has to
realize that
\begin{equation}
f(0)\equiv\varepsilon(\mathcal{W}(f))=\left.  f(x^{i})\right\vert _{x^{i}=0}.
\end{equation}
Furthermore, we took the convention that%
\begin{equation}
f(x^{i}\oplus_{A}x^{j})\equiv\overset{x|y}{\circledast}\left.  f(x^{i}%
\oplus_{A}y^{j})\right\vert _{y\rightarrow x}=f_{(1)}\circledast f_{(2)},
\end{equation}
i.e. tensor factors which are addressed by the same coordinates have to be
multiplied via the star product. The relations in (\ref{HopfAxiom}) can be
interpreted as follows. The first relation is nothing else than the law of
associativity, whereas the second and third relation concern the existence of
the identity and that of the inverse, respectively. With these rules at hand
we can now perform calculations like the following one:%
\begin{align}
f((x^{i}\oplus_{A}y^{j})\oplus_{A}(\ominus_{A}\,y^{k}))  &  =f(x^{i}\oplus
_{A}(y^{j}\oplus_{A}(\ominus_{A}\,y^{k})))\nonumber\\
&  =f(x^{i}\oplus_{A}0)=f(x^{i}).
\end{align}

As remarked in Sec. \ref{MinSpace}, q-deformed exponentials can be used to
generate q-deformed translations. The correspondence between q-exponentials
and q-translations is given by%
\begin{align}
\exp(x^{i}|\partial^{j})_{\bar{R},L}\overset{\partial|y}{\triangleright
}g(y^{k})  &  =g(x^{i}\oplus_{\bar{L}}y^{k}),\nonumber\\
\exp(x^{i}|\hat{\partial}^{j})_{R,\bar{L}}\,\overset{\partial|y}%
{\bar{\triangleright}}\,g(y^{k})  &  =g(x^{i}\oplus_{L}y^{k}),
\label{q-TayRec}%
\end{align}
and%
\begin{align}
g(y^{k})\,\overset{y|\partial}{\bar{\triangleleft}}\,\exp(\partial^{j}%
|x^{i})_{\bar{R},L}  &  =g(y^{k}\oplus_{R}x^{i}),\nonumber\\
g(y^{k})\overset{y|\partial}{\triangleleft}\exp(\hat{\partial}^{j}%
|x^{i})_{R,\bar{L}}  &  =g(y^{k}\oplus_{\bar{R}}x^{i}). \label{q-TayRecN}%
\end{align}
In the case of the quantum plane and the q-deformed Euclidean spaces these
identities can be verified in a straightforward manner. This means, we first
evaluate the left-hand side by making use of the explicit form for the
q-exponentials (see Ref. \cite{Wac03}) and the action of partial derivatives
(see Ref. \cite{BW01}). Then we compare the results with the expressions for
q-translations derived in Secs. \ref{2dimPl}-\ref{4dimEuc}.

It is very instructive to expand the q-exponentials in (\ref{q-TayRec}) and
(\ref{q-TayRecN}) up to terms linear in $x^{i}.$ In doing so, we obtain%
\begin{align}
f(x^{i}\oplus_{L}y^{j})  &  =1\otimes f(y^{j})+x^{k}\otimes\hat{\partial}%
_{k}\,\bar{\triangleright}\,f(y^{j})+O(x^{2}),\nonumber\\
f(x^{i}\oplus_{\bar{L}}y^{j})  &  =1\otimes f(y^{j})+x^{k}\otimes\partial
_{k}\triangleright f(y^{j})+O(x^{2}),\\[0.16in]
f(y^{i}\oplus_{\bar{R}}x^{j})  &  =f(y^{i})\otimes1+f(y^{i})\triangleleft
\hat{\partial}^{k}\otimes x_{k}+O(x^{2}),\nonumber\\
f(y^{i}\oplus_{R}x^{j})  &  =f(y^{i})\otimes1+f(y^{i})\,\bar{\triangleleft
}\,\partial^{k}\otimes x_{k}+O(x^{2}),
\end{align}
where repeated indices are to be summed. We see that in complete analogy to
the classical case partial derivatives generate infinitesimal translations.
This observation gives us a hint to another possibility for defining the
action of partial derivatives. From the above relations we can read off that%
\begin{align}
\partial_{i}\triangleright f(x^{j})  &  =\lim_{y^{k}\rightarrow0}\frac
{f(y^{k}\oplus_{\bar{L}}x^{j})-f(x^{j})}{y^{i}},\nonumber\\
\hat{\partial}_{i}\,\bar{\triangleright}\,f(x^{j})  &  =\lim_{y^{k}%
\rightarrow0}\frac{f(y^{k}\oplus_{L}x^{j})-f(x^{j})}{y^{i}}, \label{DiffQuo1}%
\\[0.16in]
f(x^{j})\,\bar{\triangleleft}\,\partial^{i}  &  =\lim_{y^{k}\rightarrow0}%
\frac{f(x^{j}\oplus_{R}y^{k})-f(x^{j})}{y_{i}},\nonumber\\
f(x^{j})\triangleleft\hat{\partial}^{i}  &  =\lim_{y^{k}\rightarrow0}%
\frac{f(x^{j}\oplus_{\bar{R}}y^{k})-f(x^{j})}{y_{i}}. \label{DiffQuo2}%
\end{align}

Next, we would like to illustrate that translations on quantum spaces give a
very intuitive way for describing functions being constant in the sense of
q-deformation. Such functions have to be subject to%
\begin{equation}
df=(\partial^{i}\triangleright f)dx_{i}=dx^{i}(f\triangleleft\partial_{i})=0.
\label{ConstFun1}%
\end{equation}
However, this condition is equivalent to
\begin{equation}
\partial^{i}\triangleright f=f\triangleleft\partial_{i}=0\quad\text{for
all\ }i. \label{ConstFunc2}%
\end{equation}
Notice that the same reasonings also hold for conjugate actions of partial
derivatives. A short glance at the relations in (\ref{DiffQuo1}) and
(\ref{DiffQuo2}) [or (\ref{q-TayRec}) and (\ref{q-TayRecN})] tells us that
functions being constant in the sense of (\ref{ConstFun1}) and
(\ref{ConstFunc2}) obey%
\begin{equation}
f(x^{i}\oplus_{L/\bar{L}}y^{j})=f(x^{i}),\quad f(y^{j}\oplus_{R/\bar{R}}%
x^{i})=f(x^{i}),
\end{equation}
i.e. they are invariant under q-deformed translations.

Our results about q-translations allow us to characterize functions being
invariant under translations by some sort of periodicity conditions. We wish
to illustrate the derivation of these conditions for the case of the Manin
plane. With the help of expression (\ref{CoForm2dim}) we conclude that%
\begin{align}
&  f(x^{i}\oplus_{L}y^{j})=0\nonumber\\
\Leftrightarrow\; &  D_{q^{-2}}^{1}\,f=D_{q^{-2}}^{2}\,f=0\nonumber\\
\Leftrightarrow\; &  f(q^{\pm2}x^{1})=f(q^{\pm2}x^{2})=f.
\end{align}
Proceeding in very much the same way for all quantum spaces under
consideration we obtain the following periodicity conditions for
translationally invariant functions:

\begin{enumerate}
\item (two-dimensional quantum plane)%
\begin{equation}
f(q^{\pm2}x^{1})=f(q^{2}x^{\pm2})=f, \label{PerCon1}%
\end{equation}

\item (three-dimensional q-deformed Euclidean space)%
\begin{equation}
f(q^{\pm4}x^{+})=f(q^{\pm2}x^{3})=f(q^{\pm4}x^{-})=f,
\end{equation}

\item (four-dimensional q-deformed Euclidean space)%
\begin{equation}
f(q^{\pm2}x^{1})=f(q^{\pm2}x^{2})=f(q^{\pm2}x^{3})=f(q^{\pm2}x^{4})=f,
\end{equation}

\item (q-deformed Minkowski space)%
\begin{equation}
f(q^{\pm2}r^{2})=f(q^{\pm2}x^{+})=f(q^{\pm2}x^{3/0})=f(q^{\pm2}x^{-})=f.
\label{PerCon2}%
\end{equation}

\end{enumerate}

From these periodicity conditions we can read off integrals over the whole
space, if we define an integral as a functional being invariant under
translations. To this end, we have to realize that the Jackson integral given
by \cite{Jack27}%
\begin{equation}
\left.  (D_{q^{a}})^{-1}f\right\vert _{0}^{\infty}\equiv-(1-q^{a}%
)\sum_{k=-\infty}^{\infty}(q^{-ak}x)f(q^{-ak}x),\quad q>1,
\end{equation}
satisfies%
\begin{equation}
\big ((D_{q^{a}})^{-1}f\big |_{0}^{\infty}\big )(q^{\pm a}x)=\left.
(D_{q^{a}})^{-1}f\right\vert _{0}^{\infty}.
\end{equation}
By virtue of this relation and the conditions in (\ref{PerCon1}%
)-(\ref{PerCon2}) it is not very difficult to check that the following
expressions are invariant under q-translations:

\begin{enumerate}
\item (two-dimensional quantum plane)%
\begin{equation}
\left.  (D_{q^{2}}^{1})^{-1}(D_{q^{2}}^{2})^{-1}f\right\vert _{\underline
{x}=0}^{\infty},
\end{equation}

\item (three-dimensional q-deformed Euclidean space)%
\begin{equation}
\left.  (D_{q^{4}}^{+})^{-1}(D_{q^{2}}^{3})^{-1}(D_{q^{4}}^{-})^{-1}%
f\right\vert _{\underline{x}=0}^{\infty},
\end{equation}

\item (four-dimensional q-deformed Euclidean space)%
\[
\left.  (D_{q^{2}}^{1})^{-1}(D_{q^{2}}^{2})^{-1}(D_{q^{2}}^{3})^{-1}(D_{q^{2}%
}^{4})^{-1}f\right\vert _{\underline{x}=0}^{\infty},
\]

\item (q-deformed Minkowski space)%
\begin{equation}
\left.  (D_{q^{2}}^{r^{2}})^{-1}(D_{q^{2}}^{+})^{-1}(D_{q^{2}}^{3/0}%
)^{-1}(D_{q^{2}}^{-})^{-1}f\right\vert _{\underline{x}=0}^{\infty}.
\end{equation}

\end{enumerate}

\noindent This way, we regain the q-deformed integrals we introduced in Ref.
\cite{Wac02}.

The considerations so far show us that q-deformed translations perfectly fit
into the framework of q-deformed analysis. Moreover, it should become clear
that q-deformed translations behave in a way very similar to classical
translations.\ Since translation symmetry plays a very important role in
physics, our considerations should prove useful in formulating physical
theories on quantum spaces. At this point, let us mention that in the former
literature the problem of formulating physical theories on quantum spaces is
often attacked by the construction of Hilbert space representations
\cite{CW98, Fio93, Zip95, FBM03}. Although this approach is rather rigorous
from a mathematical point of view, it is not so easy to understand the
physical meaning of its results. The reason for this lies in the fact that the
classical limit is not always quite clear. Our work, however, is much more
inspired by the so-called star-product formalism, since we consider
representations on normally ordered monomials. This approach has the great
advantage that the classical limit always exits if $q$ tends to $1.$
Unfortunately, in the case of q-deformed Minkowski space the process of normal
ordering reveals an extraordinary complexity. But our investigations also show
that in principle this complexity is manageable and with getting more
experience simplifications should be possible.\vspace{0.16in}

\noindent\textbf{Acknowledgement}\newline First of all I am very grateful to
Eberhard Zeidler for very interesting and useful discussions, special interest
in my work, and financial support. Also I wish to express my gratitude to
Julius Wess for his efforts, suggestions, and discussions. Furthermore I would
like to thank Alexander Schmidt, Fabian Bachmaier, and Florian Koch for useful
discussions and their steady support. Finally, I thank Dieter L\"{u}st for
kind hospitality.

\appendix

\section{Notation\label{AppNot}}

In this appendix we give some comments on our notation.

\begin{enumerate}
\item First of all, the antisymmetric \textit{q-number} is defined by
\cite{KS97}
\[
\left[  \left[  c\right]  \right]  _{q^{a}}\equiv\frac{1-q^{ac}}{1-q^{a}%
},\qquad a,c\in\mathbb{C}.
\]

\item For $m\in\mathbb{N}$, we can then introduce the \textit{q-factorial }by
setting
\begin{equation}
\left[  \left[  m\right]  \right]  _{q^{a}}!\equiv\left[  \left[  1\right]
\right]  _{q^{a}}\left[  \left[  2\right]  \right]  _{q^{a}}\ldots\left[
\left[  m\right]  \right]  _{q^{a}},\qquad\left[  \left[  0\right]  \right]
_{q^{a}}!\equiv1.
\end{equation}

\item There is also a q-analog of the usual binomial coefficient, the
so-called \textit{q-binomial coefficient} defined by the formula
\begin{equation}%
\genfrac{[}{]}{0pt}{}{\alpha}{m}%
_{q^{a}}\equiv\frac{\left[  \left[  \alpha\right]  \right]  _{q^{a}}\left[
\left[  \alpha-1\right]  \right]  _{q^{a}}\ldots\left[  \left[  \alpha
-m+1\right]  \right]  _{q^{a}}}{\left[  \left[  m\right]  \right]  _{q^{a}}!}%
\end{equation}
where $\alpha\in\mathbb{C},$ $m\in\mathbb{N}$.

\item The \textit{Jackson derivative} referring to the coordinate $x^{A}$ is
defined by \cite{Kac00, Jac08}%
\begin{equation}
D_{q^{a}}^{A}f\equiv\frac{f(x^{A})-f(q^{a}x^{A})}{(1-q^{a})x^{A}}%
\end{equation}
where $f$ may depend on other coordinates as well. Higher Jackson derivatives
are obtained by applying the above operator $D_{q^{a}}^{A}$ several times:
\begin{equation}
(D_{q^{a}}^{A})^{i}f\equiv\underbrace{D_{q^{a}}^{A}D_{q^{a}}^{A}\ldots
D_{q^{a}}^{A}}_{i\text{ times}}f.
\end{equation}

\item Commutative coordinates are usually denoted by small letters (e.g.
$x^{+},$ $x^{-},$ etc.), non-commutative coordinates in capital (e.g. $X^{+},$
$X^{-},$ etc.).

\item Note that in functions only such variables are explicitly displayed
which are affected by a scaling. For example, we write
\begin{equation}
f(q^{2}x^{+})\qquad\text{instead of}\qquad f(q^{2}x^{+},x^{3},x^{-}).
\end{equation}

\item Arguments enclosed in brackets shall refer to the first object on their
left. For example, we have
\begin{equation}
D_{q^{2}}^{+}f(q^{2}x^{+})=D_{q^{2}}^{+}(f(q^{2}x^{+})),
\end{equation}
or
\begin{equation}
D_{q^{2}}^{+}\big [D_{q^{2}}^{+}f+D_{q^{2}}^{-}f\big ](q^{2}x^{+})=D_{q^{2}%
}^{+}\big (\big [D_{q^{2}}^{+}f+D_{q^{2}}^{-}f\big ](q^{2}x^{+})\big ).
\end{equation}

\item Additionally, we need the operators%
\begin{equation}
\hat{n}_{A}\equiv x^{A}\frac{\partial}{\partial x^{A}}.
\end{equation}

\end{enumerate}

\section{Quantum spaces\label{AppQuan}}

The coordinates of the two-dimensional q-deformed quantum plane fulfill the
relation \cite{Man88,SS90}
\begin{equation}
X^{1}X^{2}=qX^{2}X^{1}, \label{2dimQuan}%
\end{equation}
whereas the quantum metric is given by a matrix $\varepsilon^{ij}$ with
non-vanishing elements
\begin{equation}
\varepsilon^{12}=q^{-1/2},\quad\varepsilon^{21}=-q^{1/2}.
\end{equation}
The inverse of $\varepsilon^{ij}$ is given by
\begin{equation}
(\varepsilon^{-1})^{ij}=\varepsilon_{ij}=-\varepsilon^{ij}.
\end{equation}

In the case of the q-deformed Euclidean space in three dimensions the
commutation relations read \cite{LWW97}
\begin{align}
X^{3}X^{+}  &  =q^{2}X^{+}X^{3},\nonumber\\
X^{-}X^{3}  &  =q^{2}X^{3}X^{-},\nonumber\\
X^{-}X^{+}  &  =X^{+}X^{-}+\lambda X^{3}X^{3}. \label{Koord3dimN}%
\end{align}
The non-vanishing elements of the quantum metric are
\begin{equation}
g^{+-}=-q,\quad g^{33}=1,\quad g^{-+}=-q^{-1}.
\end{equation}
Its inverse is determined by
\begin{equation}
(g^{-1})^{AB}=g_{AB}=g^{AB}.
\end{equation}

For the four-dimensional q-deformed\ Euclidean space we have the relations
\cite{CSSW90, Oca96}
\begin{align}
X^{1}X^{2}  &  =qX^{2}X^{1},\nonumber\\
X^{1}X^{3}  &  =qX^{3}X^{1},\nonumber\\
X^{3}X^{4}  &  =qX^{4}X^{3},\nonumber\\
X^{2}X^{4}  &  =qX^{4}X^{2},\nonumber\\
X^{2}X^{3}  &  =X^{3}X^{2},\nonumber\\
X^{4}X^{1}  &  =X^{1}X^{4}+\lambda X^{2}X^{3}. \label{Algebra4N}%
\end{align}
The metric has as non-vanishing components
\begin{equation}
g^{14}=q^{-1},\quad g^{23}=g^{32}=1,\quad g^{41}=q,
\end{equation}
and it holds
\begin{equation}
(g^{-1})^{ij}=g_{ij}=g^{ij}. \label{Metrik}%
\end{equation}

In the case of the q-deformed Minkowski space the coordinates obey the
relations \cite{CSSW90}
\begin{align}
X^{\mu}X^{0}  &  =X^{0}X^{\mu},\quad\mu\in\{0,+,-,3\},\nonumber\\
X^{-}X^{3}-q^{2}X^{3}X^{-}  &  =-q\lambda X^{0}X^{-},\nonumber\\
X^{3}X^{+}-q^{2}X^{+}X^{3}  &  =-q\lambda X^{0}X^{+},\nonumber\\
X^{-}X^{+}-X^{+}X^{-}  &  =\lambda(X^{3}X^{3}-X^{0}X^{3}), \label{MinrelN}%
\end{align}
and the non-vanishing components of the metric read
\begin{equation}
\eta^{00}=-1,\quad\eta^{33}=1,\quad\eta^{+-}=-q,\quad\eta^{-+}=-q^{-1}.
\end{equation}
Again, we have
\begin{equation}
(\eta^{-1})^{\mu\nu}=\eta_{\mu\nu}=\eta_{\mu\nu}.
\end{equation}
For other deformations of Minkowski spacetime we refer to Refs. \cite{Lu92,
Cas93, Dob94, ChDe95, DFR95, ChKu04, Koch04}.

\section{Coordinate Transformations\label{AppTrans}}

In our previous work\ about q-deformed Minkowski space \cite{WW01, BW01} the
algebra isomorphism $\mathcal{W}$ [cf. Eq. (\ref{AlgIsoN})] was determined by%
\begin{equation}
\mathcal{W}((x^{+})^{n_{+}}(x^{3/0})^{n_{3/0}}(x^{0})^{n_{0}}(x^{-})^{n_{-}%
})=(X^{+})^{n_{+}}(X^{3/0})^{n_{3/0}}(X^{0})^{n_{0}}(X^{-})^{n_{-}}.
\end{equation}
For the sake of simplicity the results of Sec. \ref{MinSpace} refer to a
different algebra isomorphism given by (\ref{AlgIsoMin}). In order to compare
the results of this article with those from our previous work we would like to
derive a transformation that allows us to substitute the time coordinate
$X^{0}$ of q-deformed Minkowski space by the square of the Minkowski length
$\hat{r}^{2}$ and vice versa.

To this end, we first have to realize that
\begin{equation}
X^{0}=(-\lambda_{+}^{-1})\big (q\hat{r}^{2}(X^{3/0})^{-1}+q^{-1}X^{3/0}%
-q^{-1}\lambda_{+}X^{+}(X^{3/0})^{-1}X^{-}\big ),
\end{equation}
which is a direct consequence of the definition of $\hat{r}^{2}$ [cf. Eq.
(\ref{Defr2N})]. Next, we take powers of the above expression and try to
rewrite them in terms of the monomials defining (\ref{AlgIsoMin}). With the
same reasonings already applied in Ref. \cite{WW01} [cf. Eq. (4.44)] we can
show that
\begin{equation}
(X^{0})^{n_{0}}=\sum_{p=0}^{n_{0}}q^{-p}(X^{+})^{p}(X^{3/0})^{-p}%
\,(S_{1})_{n_{0},p}(\hat{r}^{2},X^{3/0})\,(X^{-})^{p},
\end{equation}
which, in turn, leads to%
\begin{align}
(X^{+})^{n_{+}}(X^{3/0})^{n_{3/0}}(X^{0})^{n_{0}}(X^{-})^{n_{-}}= &
\sum_{p=0}^{n_{0}}q^{(2n_{3/0}-1)p}(X^{+})^{n_{+}+\,p}(X^{3/0})^{n_{3/0}%
-p}\nonumber\\
&  \times\,(S_{1})_{n_{0},p}(\hat{r}^{2},X^{3/0})\,(X^{-})^{n_{-}%
+\,p}.\label{TraX0R2N}%
\end{align}
For brevity, we introduced the polynomial%
\begin{gather}
(S_{1})_{k,p}(\hat{r}^{2},X^{3/0})\equiv\nonumber\\%
\begin{cases}
\sum_{j_{1}=0}^{p}\sum_{j_{2}=0}^{j_{1}}\ldots\sum_{j_{k-p}=0}^{j_{k-p-1}%
}\prod\limits_{l=1}^{k-p}a_{1}(\hat{r}^{2},q^{2j_{l}}X^{3/0}), & \text{if
}0\leq p<k,\\
1 & \text{if }p=k,
\end{cases}
\end{gather}
with
\begin{equation}
a_{1}(\hat{r}^{2},X^{3/0})\equiv-\lambda_{+}^{-1}\big (q\hat{r}^{2}%
(X^{3/0})^{-1}+q^{-1}X^{3/0}\big ).
\end{equation}
Recalling that $\hat{r}^{2}$ is central in the Minkowski space algebra we
conclude that (\ref{TraX0R2N}) indeed allows us to substitute the time
coordinate $X^{0}$ by the square of the Minkowski length $\hat{r}^{2}$.

However, the expression in (\ref{TraX0R2N}) is not suitable to read off an
operator that performs the coordinate transformation on commutative functions.
To get round this problem, we proceed in the following way:%
\begin{align}
(X^{0})^{n_{0}}=  &  \,(-\lambda_{+}^{-1})^{n_{0}}\big (q\hat{r}^{2}%
(X^{3/0})^{-1}+q^{-1}X^{3/0}-q^{-1}\lambda_{+}X^{+}(X^{3/0})^{-1}%
X^{-}\big )^{n_{0}}\nonumber\\
=  &  \sum_{k=0}^{n_{0}}(-\lambda_{+}^{-1})^{n_{0}}\binom{n_{0}}{k}(q\hat
{r}^{2}(X^{3/0})^{-1})^{n_{0}-k}\nonumber\\
&  \,\qquad\times\big (q^{-1}X^{3/0}-q^{-1}\lambda_{+}X^{+}(X^{3/0})^{-1}%
X^{-}\big )^{k}\nonumber\\
=  &  \,\sum_{k=0}^{n_{0}}\sum_{l=0}^{k}(-1)^{n_{0}+k+l}q^{n_{0}%
-2k-l(l+1)}\lambda_{+}^{l-n_{0}}\binom{n_{0}}{k}\binom{k}{l}\nonumber\\
&  \,\qquad\times(\hat{r}^{2})^{n_{0}-k}(X^{3/0})^{-n_{0}+2(k-l)}(X^{+}%
X^{-})^{l}. \label{X0N0N}%
\end{align}
In addition to this, we have
\begin{align}
(X^{+}X^{-})^{l}=  &  \big (\lambda_{+}^{-1}\hat{r}^{2}+q^{-1}X^{0}%
X^{3/0}+q^{-2}\lambda_{+}^{-1}X^{3/0}X^{3/0}\big )^{l}\nonumber\\
=  &  \sum_{i=0}^{l}(\lambda_{+})^{-l}\binom{l}{i}(\hat{r}^{2})^{i}%
\big (a_{2}(X^{0},X^{3/0})\big )^{l-i}, \label{XPXmUmN}%
\end{align}
where
\begin{equation}
a_{2}(X^{0},X^{3/0})\equiv q^{-2}(X^{3/0})^{2}+q^{-1}\lambda_{+}X^{0}X^{3/0}.
\label{ApExp}%
\end{equation}
It seems so that we did not made any progress, since the last expression in
(\ref{XPXmUmN}) still depends on $X^{0}$. However, we are able to rewrite
powers of the expression in Eqn. (\ref{ApExp}) as follows:
\begin{align}
&  \hspace{-0.16in}\big (a_{2}(X^{0},X^{3/0})\big )^{m}=\big (q^{-2}%
(X^{3/0})^{2}+q^{-1}\lambda_{+}X^{0}X^{3/0}\big )^{m}\nonumber\\
=  &  \,\sum_{u=0}^{m}(q^{-1}\lambda_{+})^{u}(q^{-2})^{m-u}\binom{m}%
{u}(X^{3/0})^{2m-u}(X^{0})^{u}\nonumber\\
=  &  \,\sum_{u=0}^{m}\sum_{v=0}^{u}q^{2m(2v-1)-u(2v-1)-v}\lambda_{+}%
^{u}\binom{m}{u}\nonumber\\
&  \,\qquad\times(X^{+})^{v}(X^{3/0})^{2m-u-v}\,(S_{1})_{u,v}(\hat{r}%
^{2},X^{3/0})\,(X^{-})^{v}.
\end{align}
Notice that the third equality holds due to Eqn. (\ref{TraX0R2N}). Putting
everything together we finally get
\begin{align}
&  \hspace{-0.16in}(X^{+})^{n_{+}}(X^{0})^{n_{0}}(X^{3/0})^{n_{3/0}}%
(X^{-})^{n_{-}}=\sum_{k=0}^{n_{0}}\sum_{l=0}^{k}\sum_{i=0}^{l}\sum
_{u=0}^{l-\,i}\sum_{v=0}^{u}(-1)^{n_{0}+k+l}\,\lambda_{+}^{-n_{0}%
+u}\nonumber\\
&  \qquad\qquad\times q^{-l(l-1)-u(2v-1)+v(4(k-i)-1)-2(l+k-i)}\nonumber\\
&  \qquad\qquad\times q^{2v(n_{3/0}-\,n_{0})+n_{0}}\binom{n_{0}}{k}\binom
{k}{l}\binom{l}{i}\binom{l-i}{u}\nonumber\\
&  \qquad\qquad\times(X^{+})^{n_{+}+\,v}(\hat{r}^{2})^{n_{0}-\,k+i}%
\,(S_{1})_{u,v}(\hat{r}^{2},X^{3/0})\nonumber\\
&  \qquad\qquad\times(X^{3/0})^{n_{3/0}-\,n_{0}+2(k-i)-u-v}(X^{-})^{n_{-}%
+\,v}.
\end{align}
From the above identity we can now read off the operator
\begin{align}
&  \hspace{-0.16in}\hat{T}f(x^{0},x^{+},x^{3/0},x^{-})=\sum_{k=0}^{\infty}%
\sum_{l=0}^{k}\sum_{i=0}^{l}\sum_{u=0}^{l-\,i}\sum_{v=0}^{u}(-1)^{k+l}%
\frac{\lambda_{+}^{u}}{k!}\nonumber\\
&  \qquad\times q^{-l(l+1)-u(2v-1)+v(4(k-i)-1)-2(l+k-i)}\binom{k}{l}\binom
{l}{i}\binom{l-i}{u}\nonumber\\
&  \qquad\times(x^{+}x^{-})^{v}(x^{3/0})^{k-2i-u-v}(r^{2})^{i}\,(S_{1}%
)_{u,v}(r^{2},x^{3/0})\nonumber\\
&  \qquad\times\big ((\partial/\partial x^{0})^{k}f(-q^{1-2v}\lambda_{+}%
^{-1}x^{0},q^{2v}x^{3/0})\big )\big |_{x^{0}\rightarrow\hat{r}^{2}%
(x^{3/0})^{-1}},
\end{align}
where $\partial/\partial x^{0}$ denotes the partial derivative for the
commutative coordinate $x^{0}.$

It remains to find the operator for the inverse coordinate transformation. In
very much the same way as above we can derive from the definition of $\hat
{r}^{2}$ the identity
\begin{align}
\hat{r}^{2n}=  &  (\lambda_{+}X^{+}X^{-}-q^{-1}\lambda_{+}X^{0}X^{3/0}%
-q^{-2}X^{3/0}X^{3/0})^{n}\nonumber\\
=  &  \sum_{k=0}^{n}\sum_{l=0}^{n-k}\sum_{i=0}^{k}\sum_{p=0}^{i}%
(-1)^{n-k}q^{(n-k+l)(2p-1)}(\lambda_{+})^{p+n-k-l}\nonumber\\
&  \qquad\times\binom{n}{k}\binom{n-k}{l}\binom{k}{i}(X^{+})^{p}%
(X^{3/0})^{n-k+l}(X^{0})^{n-k-l}\nonumber\\
&  \qquad\times\big (a_{2}(X^{0},q^{2p}X^{3/0})\big )^{k-i}\,(S_{2}%
)_{i,p}(X^{0},X^{3/0})\,(X^{-})^{p},
\end{align}
leading us to the inverse of $\hat{T}$, i.e.
\begin{align}
&  \hspace{-0.16in}\hat{T}^{-1}f(r^{2},x^{+},x^{3/0},x^{-})=\sum_{k=0}%
^{\infty}\sum_{l=0}^{\infty}\sum_{i=0}^{k}\sum_{p=0}^{i}(-1)^{k}%
q^{(2p-1)(l-k)}\frac{\lambda_{+}^{p}}{k!\,l!}\binom{k}{i}\nonumber\\
&  \qquad\times(x^{+}x^{-})^{p}(x^{3/0})^{2l}\big (a_{2}(x^{0},q^{2p}%
x^{3/0})\big )^{k-i}\,(S_{2})_{i,p}(x^{0},x^{3/0})\nonumber\\
&  \qquad\times\left.  \big ((\partial/\partial r^{2})^{k+l}f(q^{2p}%
x^{3/0},-q^{2p-1}\lambda_{+}r^{2})\big )\right\vert _{r^{2}\rightarrow
x^{0}x^{3/0}},
\end{align}
where
\begin{gather}
(S_{2})_{k,p}(x^{0},x^{3/0})\equiv\nonumber\\%
\begin{cases}
-\sum_{j_{1=0}}^{p}\sum_{j_{2=0}}^{j_{1}}\ldots\sum_{j_{k-p=0}}^{j_{k-p-1}%
}\prod_{l=1}^{k-p}a_{2}(x^{0},q^{2j_{l}}x^{3/0}), & \text{if }0\leq p<k,\\
1, & \text{if }p=k,
\end{cases}
\label{FlendN}%
\end{gather}
and $\partial/\partial r^{2}$ denotes the partial derivative for the
commutative variable $r^{2}.$

\end{document}